\documentclass[twocolumn]{aastex63}

\usepackage{textcomp}
\usepackage{comment}
\usepackage{lineno}
\usepackage{amsmath}
\newcommand{\mbh}{\ifmmode M_{\rm BH} \else $M_{\rm BH}$\fi}

\shorttitle{Disk emitters in BASS}
\shortauthors{Ward et al.}

\graphicspath{{./}{figures/}}

\begin{document}

\title{BASS LII: The prevalence of double-peaked broad lines at low accretion rates \\among hard X-ray selected AGN}
\correspondingauthor{Charlotte Ward}
\email{cvw5890@psu.edu}

\author[0000-0002-4557-6682]{Charlotte Ward}
\affil{Department of Astronomy \& Astrophysics, 525 Davey Lab, 251 Pollock Road, The Pennsylvania State University, University Park, PA 16802, USA} 

\author[0000-0002-7998-9581]{Michael J. Koss}
\affiliation{Eureka Scientific, 2452 Delmer Street, Suite 100, Oakland, CA 94602-3017, USA}
\affiliation{Space Science Institute, 4750 Walnut Street, Suite 205, Boulder, CO 80301, USA}

\author[0000-0002-3719-940X]{Michael Eracleous}
\affiliation{Department of Astronomy \& Astrophysics and Institute for Gravitation and the Cosmos, 525 Davey Lab, 251 Pollock Road, The Pennsylvania State University, University Park, PA 16802, USA}

\author[0000-0002-3683-7297]{Benny Trakhtenbrot}
\affiliation{School of Physics and Astronomy, Tel Aviv University, Tel Aviv 69978, Israel}


\author[0000-0002-8686-8737]{Franz E. Bauer}
\affiliation{Instituto de Alta Investigaci{\'{o}}n, Universidad de Tarapac{\'{a}}, Casilla 7D, Arica, Chile}

\author[0000-0002-9144-2255]{Turgay Caglar}
\affiliation{George P. and Cynthia Woods Mitchell Institute for Fundamental Physics and Astronomy, Texas A\&M University, College Station, TX, 77845, USA}
\affiliation{Leiden Observatory, PO Box 9513, 2300 RA Leiden, The Netherlands}

\author[0000-0002-4226-8959]{Fiona Harrison}
\affiliation{Cahill Center for Astronomy and Astrophysics, California Institute of Technology, Pasadena, CA 91125, USA}

\author[0000-0001-7500-5752]{Arghajit Jana}
\affiliation{Instituto de Estudios Astrof\'isicos, Facultad de Ingenier\'ia y Ciencias, Universidad Diego Portales, Av. Ej\'ercito Libertador 441, Santiago, Chile}

\author[0000-0002-2603-2639]{Darshan Kakkad}
\affiliation{
Centre for Astrophysics Research, University of Hertfordshire, College Lane, Hatfield, AL10 9AB, UK}

\author[0000-0002-1292-1451]{Macon Magno}
\affiliation{George P. and Cynthia Woods Mitchell Institute for Fundamental Physics and Astronomy, Texas A\&M University, College Station, TX, 77845, USA}
\affiliation{CSIRO Space and Astronomy, ATNF, PO Box 1130, Bentley WA 6102, Australia}

\author[0000-0001-8931-1152]{Ignacio del Moral-Castro}
\affiliation{Instituto de Astrof\'isica, Facultad de F\'isica, Pontificia Universidad Cat\'olica de Chile, Av. Vicu\~na Mackenna 4860, Santiago, Chile}

\author[0000-0002-7962-5446]{Richard Mushotzky}
\affiliation{Department of Astronomy, University of Maryland, College Park, MD 20742, USA}
\affiliation{Joint Space-Science Institute, University of Maryland, College Park, MD 20742, USA}

\author[0000-0002-5037-951X]{Kyuseok Oh}
\affiliation{Korea Astronomy and Space Science Institute, Daedeokdae-ro 776, Yuseong-gu, Daejeon 34055, Republic of Korea}

\author[0000-0003-2196-3298]{Alessandro Peca}
\affiliation{Eureka Scientific, 2452 Delmer Street, Suite 100, Oakland, CA 94602-3017, USA}
\affiliation{Department of Physics, Yale University, P.O. Box 208120, New Haven, CT 06520, USA}

\author[0000-0003-2284-8603]{Meredith C. Powell}
\affiliation{Leibniz-Institut f\"ur Astrophysik Potsdam (AIP), An der Sternwarte 16, 14482 Potsdam, Germany}

\author[0000-0001-5231-2645]{Claudio Ricci}
\affiliation{Instituto de Estudios Astrof\'isicos, Facultad de Ingenier\'ia y Ciencias, Universidad Diego Portales, Av. Ej\'ercito Libertador 441, Santiago, Chile}
\affiliation{Kavli Institute for Astronomy and Astrophysics, Peking University, Beijing 100871, China}

\author[0000-0003-0006-8681]{Alejandra Rojas}
\affiliation{Departamento de F\'isica, Universidad T\'ecnica Federico Santa Mar\'ia, Vicu\~{n}a Mackenna 3939, San Joaqu\'in, Santiago de Chile, Chile}

\author[0000-0001-5785-7038]{Krista Lynne Smith}
\affiliation{George P. and Cynthia Woods Mitchell Institute for Fundamental Physics and Astronomy\\ Texas A\&M University\\ College Station, TX 77843-4242, USA}

\author[0000-0003-2686-9241]{Daniel Stern}
\affiliation{Jet Propulsion Laboratory, California Institute of Technology, 4800 Oak Grove Drive, MS 169-224, Pasadena, CA 91109, USA}

\author[0000-0001-7568-6412]{Ezequiel Treister}
\affiliation{Instituto de Alta Investigaci{\'{o}}n, Universidad de Tarapac{\'{a}}, Casilla 7D, Arica, Chile}

\author[0000-0002-0745-9792]{C. Megan Urry}
\affiliation{Yale Center for Astronomy \& Astrophysics and Department of Physics, Yale University, P.O. Box 208120, New Haven, CT 06520-8120, USA}
\affiliation{Department of Physics, Yale University, P.O. Box 208120, New Haven, CT 06520, USA}

\begin{abstract}
A fraction of active galactic nuclei (AGN) have double-peaked H$\alpha$, H$\beta$ and Mg II broad lines attributed to emission from rotating gas in the accretion disk. Using optical spectroscopy of a flux-limited sample of AGN selected via ultrahard X-rays from the BAT AGN Spectroscopic Survey (BASS), we systematically identify 71 double-peaked emitters amongst 343 broad-line AGN with redshifts $0.004<z<0.297$ and X-ray luminosities of $40<\log L_{2-10\text{KeV}}$(erg s$^{-1}$)$<45.7$, and provide their best-fit accretion disk parameters. We find that $\sim21$\% of X-ray selected broad-line AGN are double-peaked emitters (DPEs), consistent with rates previously reported for $z<0.2$ broad-line AGN selected for strong optical variability in ZTF. 11 of 71 DPEs (15\%) exhibited a single-peaked Gaussian component to the broad line profile in addition to the double-peaked disk profile. In this sample, DPEs have intrinsically higher masses by $\sim0.4$ dex and lower Eddington ratios by $\sim0.3$ dex than other broad-line AGN, and have a preference for elliptical host galaxies, higher X-ray luminosities, and higher [O\,{\sc i}] $\lambda$6302 to narrow H$\alpha$ flux ratios than other broad-line AGN. We find that DPEs are not segregated from non-DPE broad-line AGN in the $L_{\text{bol}}$ vs $M_{\text{BH}}$ relation or their X-ray to radio luminosity ratios, and do not show a preference for intermediate Seyfert types over Seyfert 1s. We do not find differences in a wide range of multi-wavelength properties when comparing DPEs to non-DPE broad-line AGN, including optical and mid-IR variability levels, WISE colors, $\alpha_{\text{ox}}$, the column density of neutral obscuring material $N_H$, and the rate of changing-look events. We discuss the two populations in the context of multi-component disk-wind models of the AGN broad line region and consider how unrecognized contributions of disk emission to the broad lines may introduce  biases in `virial' SMBH mass estimates, with consequences for the inferred $M_{\text{BH}}-M_{*}$ relation.

\vspace{1cm}
\end{abstract}

\section{Introduction}

Understanding how today's supermassive black holes (SMBHs) first formed, grew over time, and co-evolved with their host galaxies is key to many open questions in modern astrophysics \citep[e.g.,][]{Greene2020Intermediate-MassHoles}. Yet, we still do not understand the geometry of accretion flows and `broad line region' (BLR) gas within thousands of gravitational radii of active galactic nuclei (AGN). It is known from reverberation mapping campaigns that the motion of the BLR is not predominantly radial \citep[see][for a review]{Peterson1993} and that the BLR is virialized \citep{Peterson2000,Kollatschny2003}. A range of models for the BLR have invoked the outer part of the accretion disk as a key component, including models of clouds flowing away from the disk along magnetic field lines \citep{Emmering1992}, radiatively accelerated winds \citep{Murray1995}, and combined contributions of an accretion disk and an additional gaseous region \citep{Popovic2004ContributionModel}. Yet, several mysteries, including how super-Eddington accretion might grow the over-massive SMBHs discovered by the \textit{James Webb Space Telescope} at $z\approx10$ \citep{Goulding2023}, and how AGN broad Balmer emission lines appear and disappear on human timescales in `changing-look' events \citep{Ricci2023Changing-lookNuclei}, motivate us to understand how BLR structure depends on accretion rate, and how viewing angle affects the observable properties of broad-line AGN populations across redshift. 

A key sub-population of AGN that can provide constraints for various BLR models are the double-peaked emitters (DPEs). DPEs are typically observed to have H$\alpha$, H$\beta$ and Mg II $\lambda$2798 broad lines with two shoulders that are blueshifted and redshifted from the rest velocity by a few hundred to a few thousand km s$^{-1}$ due to the Keplerian motion of gas within a disk spanning a few hundred to a few thousand gravitational radii \citep{Eracleous1994Doubled-peakedNuclei,Chen1989, Eracleous1997RejectionNucleib,Strateva2003Double-peakedNuclei,Eracleous2009Double-peakedNuclei}. Models applied to fit the broad-line profile shapes of DPEs depend on factors such as disk inclination angle, turbulent broadening, the disk emissivity as a function of radius, and radiative transfer through the base of a wind \citep{Chen1989,Strateva2003Double-peakedNuclei,Murray1996Wind-dominatedStars,Flohic2012EFFECTSNUCLEI, Nguyen2018EmissionLines, Chajet2013MagnetohydrodynamicDistributions}. While the Balmer series and the Mg II $\lambda$2798 lines in DPEs usually have double-peaked profiles that can be described by the same disk model for a given source, UV resonance lines, such as Ly$\alpha$ and C IV, which have much higher optical depth, have single-peaked profiles due to their higher optical depth and, in the case of Ly$\alpha$, collisional de-excitation in the dense emission region \citep[][and references therein]{Eracleous2009Double-peakedNuclei}. While the binary SMBH interpretation is sometimes invoked for double-peaked profiles, this model cannot explain line profile variability observed in the majority of DPEs \citep[][and references therein]{Doan2020AnLines} and has been ruled out for some DPE samples by the observation of single-peaked Ly$\alpha$ broad-lines \citep{Runnoe2025}.

DPEs are most commonly identified in low luminosity, low-accretion rate AGN \citep{Eracleous1994Doubled-peakedNuclei,Ho2000DoublepeakedLINERs, Eracleous2003CompletionNuclei,Ho2008NuclearGalaxies}. A two-component disk-wind structure may account for a transition to a `disk-dominated' state at low luminosities \citep{Elitzur2009OnNuclei,Elitzur2014EvolutionNuclei}.  In this model, the BLR consists of an outflow of gas embedded in a hydromagnetic disk wind. When the gas is released from the disk, it expands and reduces in column density, resulting in a toroidal geometry for both the BLR and the dusty torus beyond the dust sublimation radius \citep{Emmering1992,Netzer1993,Elitzur2006,Czerny2015,Czerny2016}. The outflowing, low column density gas produces broad Gaussian lines, while high density gas close to the disk surface rotating at Keplerian velocities produces the double-peaked emission. As the accretion rate decreases, the BLR structure loses its components at the highest elevations above the disk and the double-peaked emission from the disk could increasingly dominate over Gaussian broad line emission. This model may also explain the DPE preference for intermediate-type Seyferts over Seyfert 1s \citep{Elitzur2009OnNuclei,Elitzur2014EvolutionNuclei}. In order to test this model, we need to disentangle viewing angle effects, accretion rate effects, and AGN population selection biases, motivating the systematic classification of DPE sub-samples amongst AGN selected by different means. 

Distinguishing between intrinsic differences in BLR geometry and viewing angle effects is challenging. \citet{Storchi-Bergmann2017Double-PeakedNuclei} consider that if double-peaked profiles are ubiquitous in broad-line AGN, but are usually only observed when the inclination angle is $\gtrsim20$\textdegree \ so that the separate peaks of the accretion disk are observable, but $\lesssim37$\textdegree \ so that the accretion disk emission is not blocked by the obscuring torus, this would result in an observed $\sim60$\% fraction of broad-line AGN with double peaks, which is much higher than the observed DPE rate. It is clear that outflowing gas must dominate broad-line emission in most AGN even if an additional disk emission line is present. It is known that some AGN have both a Gaussian BLR component and a disk component. This has been identified in SDSS J125809.31+351943.0, for example, which varied in Eddington ratio from 0.4\% to 2.4\%, and exhibited a primarily double-peaked H$\beta$ profile during dim periods and a single-peaked profile during bright periods \citep{Nagoshi2024}. Analysis of the `root mean square' spectra derived from multi-epoch observations of some AGN has also often shown a double-peaked H$\beta$ profile implying the presence of variable, disk-like gas that may not be clear from a single-epoch spectrum that appears single-peaked \citep{Denney2010ReverberationGalaxies, Schimoia2017,Storchi-Bergmann2017Double-PeakedNuclei}.

DPEs may be an important source of population-level bias in virial black hole mass ($M_{\text{BH}}$) measurements \citep{Bian2007ActiveRatios,Lewis2006BlackLines,Wu2004BlackLines,Fu2023}. A key method to measure $M_{\text{BH}}$ for unobscured AGN is the `virial' method, whereby the width of the H$\alpha$, H$\beta$ or other broad emission lines is used as a proxy for the virial velocity of the broad line region (BLR) gas \citep[e.g.][and references therein]{Greene2005}. By adopting the empirical relation between the BLR size and the continuum luminosity and the emission-line width of the broad component, black hole masses can be calculated \citep{Kaspi2000, Bentz2009}. While this method has been effective in determining the black hole mass function and probing the relationship between $M_{\text{BH}}$ and the stellar velocity dispersion of the host galaxy (the $M_{\text{BH}}-\sigma_{*}$ relation) for populations of unobscured AGN over a range of redshifts \citep[e.g.][]{Shen2008,Shen2019,Ubler2023,Mezcua2024}, it is sensitive to various biases that may affect $M_{\text{BH}}$ measurements by $\gtrsim0.4$ dex \citep{Shen2013, Peterson2014,Caglar2020}, including the virial factor $f$ which introduces uncertainties of 0.12 dex \citep{Woo2015}, intrinsic scatter of BLR radius-luminosity relation which introduces scatter of order 0.1 dex \citep{Bentz2009}, and variability in both the AGN continuum luminosity and the broad line width \citep{Park2012}. Double-peaked emitters have been identified in various studies where they have also inhibited accurate virial mass measurements for those objects, including in samples of hard X-ray selected AGN in the local Universe, where 29 out of $\sim600$ AGN were identified as DPE candidates \citep{Mejia2022}, as well as in \textit{James Webb Space Telescope} studies of $z>6$ AGN, where one in a sample of 12 was identified as a DPE \citep{Onoue2024}. It is therefore important to understand the fraction of broad-line AGN that are DPEs, methods to identify them, and the extent to which they may introduce biases or scatter in the $M_{\text{BH}}-\sigma_{*}$ relation.

Estimates of DPE fractions among the wider broad-line AGN population range from $\sim 3-30$\% \citep{Eracleous1994Doubled-peakedNuclei,Ho1997AEmission,Strateva2003Double-peakedNuclei,Ward2024}, depending on how the AGN sample is selected, and the criteria used to distinguish between DPEs and non-DPE broad-line AGN. Notably, the DPE rate was found to be $19\%$ amongst $z<0.4$ AGN with strong ($>$ 1.5 mag) optical variability in Zwicky Transient Facility \citep[ZTF;][]{Bellm2019,Graham2019,Dekany2020TheSystem} photometry \citep{Ward2024}. Comparison of ZTF light curve power spectra from DPEs and other variable AGN did not find significant differences in variability amplitude or power law index \citep{Ward2022Variability-selectedWISE}, despite previous work from \citet{Zhang2017PropertiesLines} finding that DPE light curves from the Catalina Sky Survey \citep{Drake2009} had damped random walk (DRW) characteristic timescales 2.7 times longer than a control sample's SDSS Stripe 82 \citep{Bramich2008} light curves. DPEs show some clearer differences to other broad-line AGN when considering their host galaxies and X-ray/radio properties: DPEs are preferentially associated with radio-loud elliptical hosts with large bulge and black hole masses \citep{Eracleous1994Doubled-peakedNuclei,Eracleous2003CompletionNuclei}, and DPE radio and soft X-ray luminosities are $\sim1.5$ times higher than control samples \citep{Strateva2003Double-peakedNuclei}. 

The BAT AGN Spectroscopic Survey\footnote{\url{https://bass-survey.com}} (BASS) provides a comprehensive dataset of optical and NIR spectroscopy for a sample of $\sim1000$ AGN selected via ultra-hard X-rays (14–195 keV) in the Swift/BAT mission \citep{Koss2017,Koss2022a}. This well-studied sample provides an opportunity to estimate DPE rates in a hard X-ray flux-limited AGN population that is not biased by viewing angle effects, and to study their multi-wavelength properties compared to other broad-line AGN. This paper is structured as follows. Section 2 outlines our modeling of optical spectra of the BASS AGN sample to distinguish between DPEs and other broad-line AGN and characterize their disk properties. In Section 3 we discuss the construction of optical ZTF light curves and mid-IR Wide--field Infrared Survey Explorer \citep[\textit{WISE};][]{Wright2010ThePerformance} light curves and variability metrics. In Section 4, we compare various X-ray, IR, and radio characteristics of the DPEs and other broad-line AGN in the BASS sample. In Section 5, we discuss the stellar masses of the DPEs and other broad-line AGN, and discuss possible biases in the virial masses measured for DPE candidates. Section 6 discusses the host galaxy properties of the two samples. Section 7 discusses the implications of our results for the disk-wind model and possible future work. We summarize our conclusions in Section 8. Throughout the paper we adopt the WMAP7 cosmology with $H_0 = 70.4$ and $\Omega_m = 0.272$ \citep{Komatsu2011}. 

\section{Disk profile fitting for BASS broad-line AGN}

With the goal of identifying double-peaked emitters amongst BASS broad-line AGN using the H$\alpha$ line, we began with a sample of 742 hard X-ray selected AGN with rest-frame optical line measurements reported in the BASS DR2 Spectroscopic Line Measurements Catalog \citep{Oh2022}. We excluded 196 Type 2 AGNs from the sample. Following the methods described in \citet{Ward2024}, we used Penalized Pixel Fitting \citep[pPXF;][]{Cappellari2003,Cappellari2017ImprovingFunctions} to model and subtract the stellar continuum and absorption lines. After removing 179 objects that did not have H$\alpha$ line coverage in the BASS spectra or had data reduction artifacts preventing the production of a reasonable continuum model, we obtained a sample of 367 AGN. We note here that the H$\beta$ broad line can also be used to constrain disk parameters, but as the H$\alpha$ profiles had higher S/N and we wished to compare to previous work from H$\alpha$ fitting of variability-selected DPEs from ZTF \citep{Ward2024}, we decided to focus only on objects with available H$\alpha$ to ensure uniformity in the fitting procedure across samples. We further excluded 24 objects with low S/N broad lines that would prevent a meaningful disk model fit. We did not apply a particular S/N cut, but instead decided on whether there was a sufficient S/N based on the posteriors obtained from the disk model fitting described in the next paragraph - if the broad line parameters were entirely unconstrained, we removed them from the sample. This process provided a sample of 343 broad-line AGN for H$\alpha$ fitting. The final sample of 343 AGN is representative of BASS unobscured AGN within a redshift range of 0.004 to 0.3. Using the galaxy morphology classifications reported in \citet{ParraTello2025}, 116 of the host galaxies were reported as having `smooth' (elliptical) morphologies, 44 are mergers, 97 are `disk' galaxies, 16 were edge-on galaxies, 29 were `point-like', and 38 fell into the `uncertain/other' category.

\begin{figure*}
\gridline{\fig{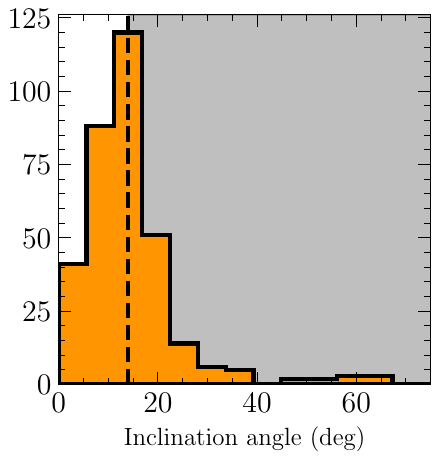}{0.32\textwidth}{} \fig{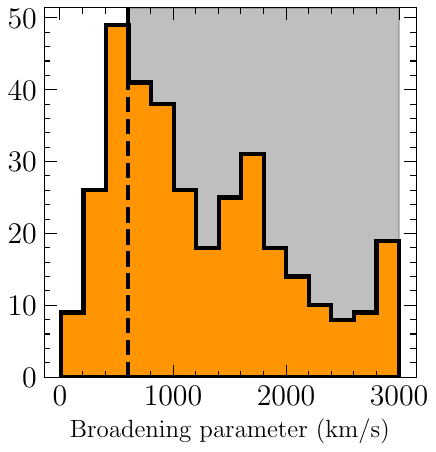}{0.33\textwidth}{} \fig{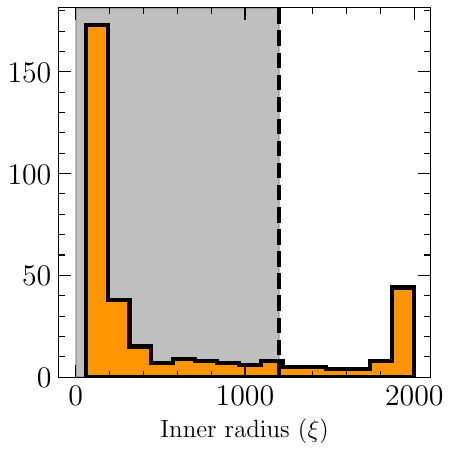}{0.33\textwidth}{}}
\caption{Histograms showing the distributions of the three disk parameters used to separate DPEs from AGN without double-peaked broad lines. DPEs were classified as those with inclination angle $>14$ \textdegree, turbulent broadening $>600$ km s$^{-1}$, and inner radius $<1200$ gravitational radii, corresponding to the gray shaded regions to the right of the vertical dashed lines in the left and center plots, and to the left of the vertical dashed line in the right plot.}
\label{fig:disksplit}
\end{figure*}

\begin{figure*}
\gridline{\fig{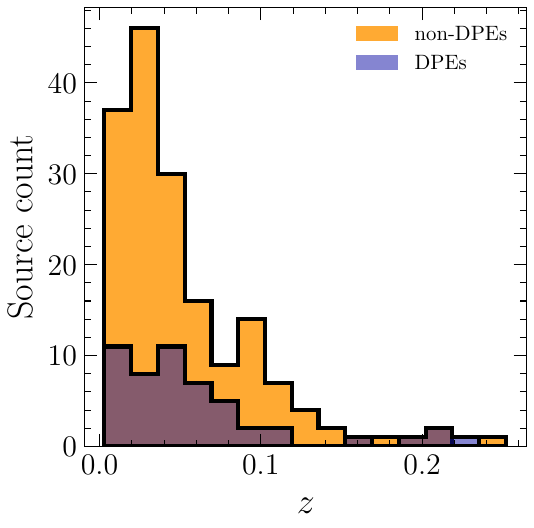}{0.32\textwidth}{} \fig{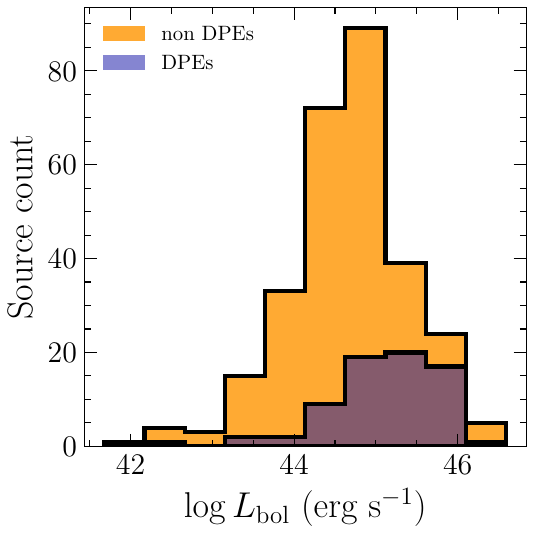}{0.32\textwidth}{} \fig{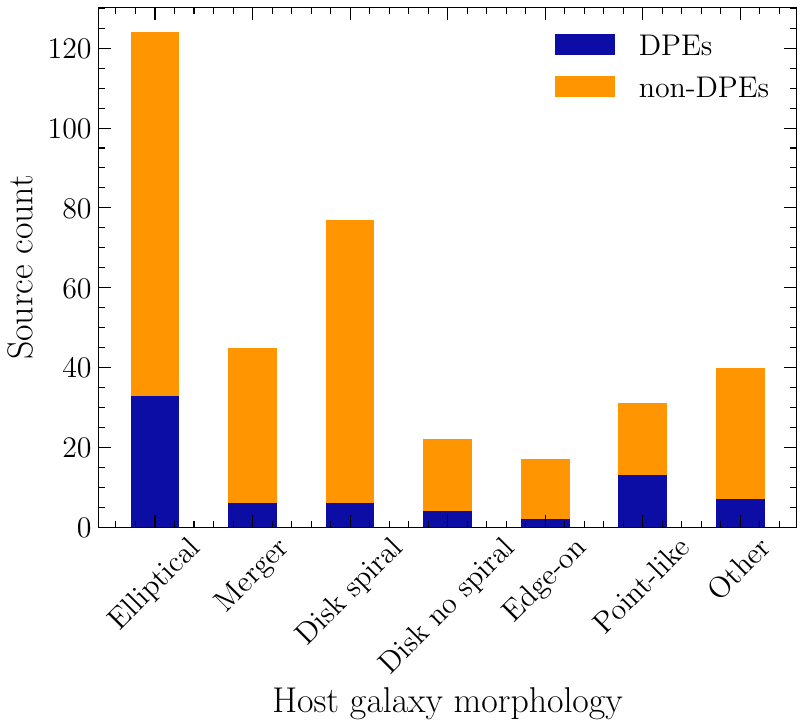}{0.35\textwidth}{}}
\caption{Properties of the parent sample after splitting into DPEs and non-DPE AGN: redshift (left), bolometric luminosity derived from the 14-150 keV intrinsic luminosities (center), and host galaxy types according to the classifications of \citet{ParraTello2025} (right).}
\label{fig:parentsample}
\end{figure*}

\begin{figure*}
\gridline{\fig{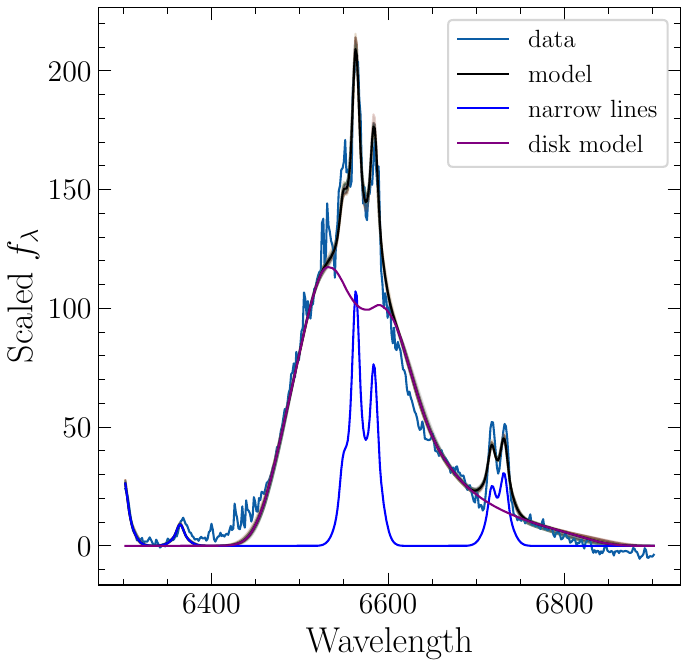}{0.24\textwidth}{a) BAT 61}\fig{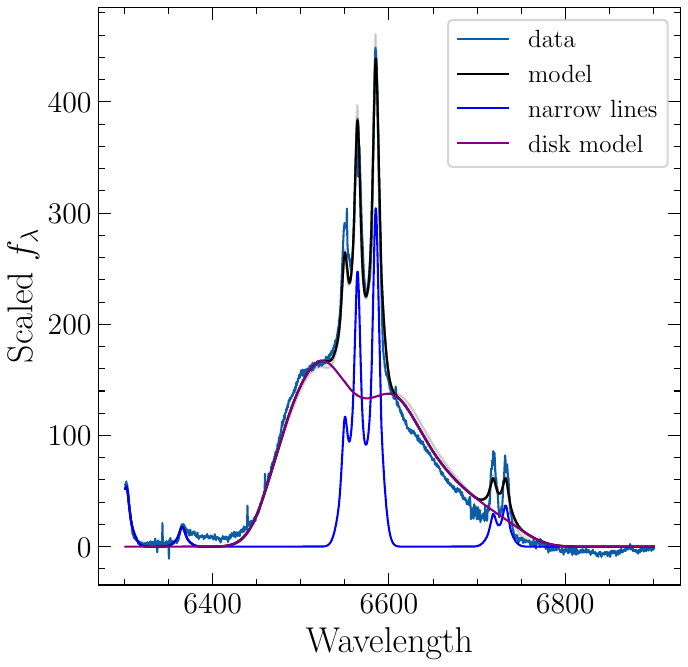}{0.24\textwidth}{b) BAT 116}\fig{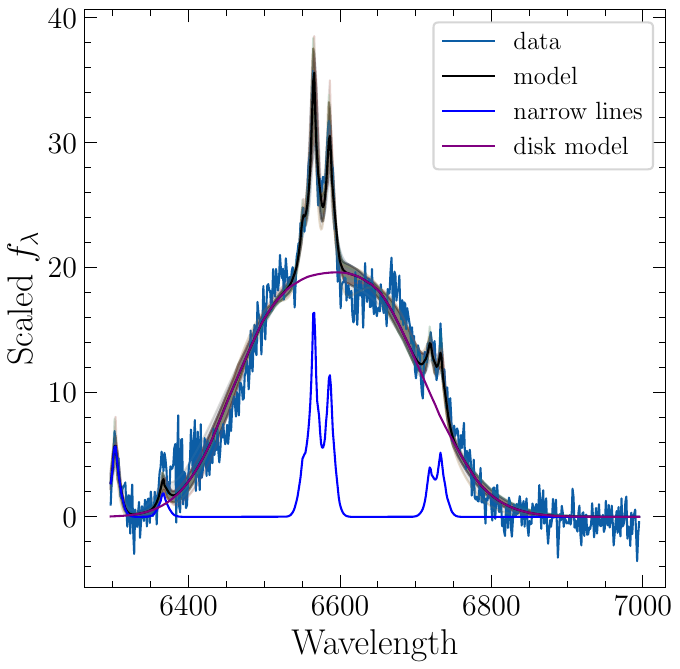}{0.24\textwidth}{c) BAT 371} \fig{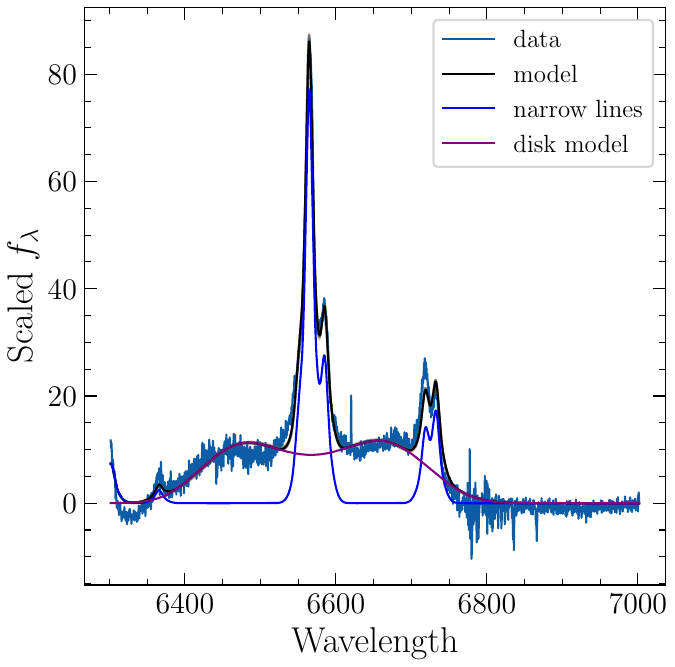}{0.24\textwidth}{d) BAT 372}}
\gridline{\fig{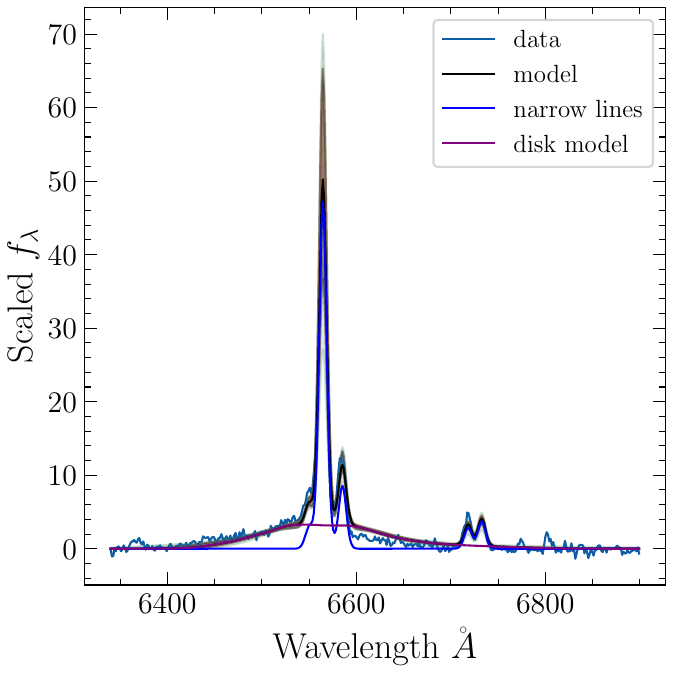}{0.24\textwidth}{e) BAT 381} 
\fig{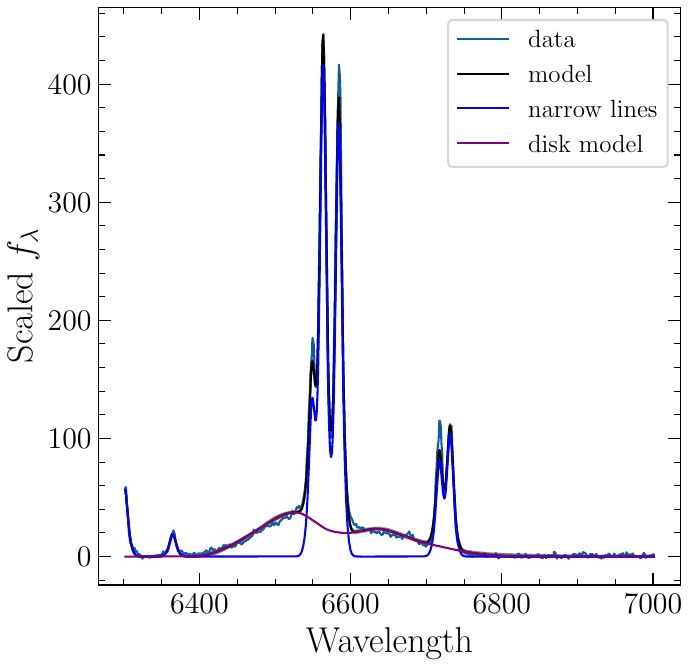}{0.24\textwidth}{f) BAT 757}\fig{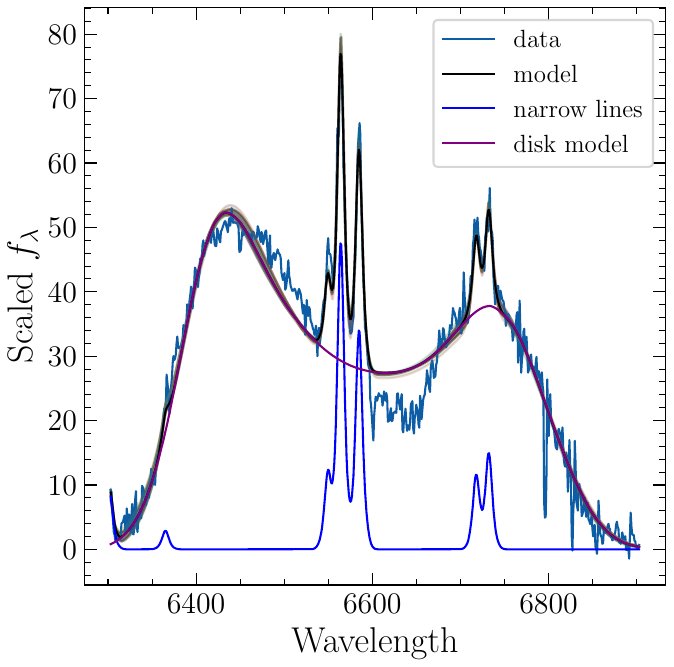}{0.24\textwidth}{g) BAT 800}\fig{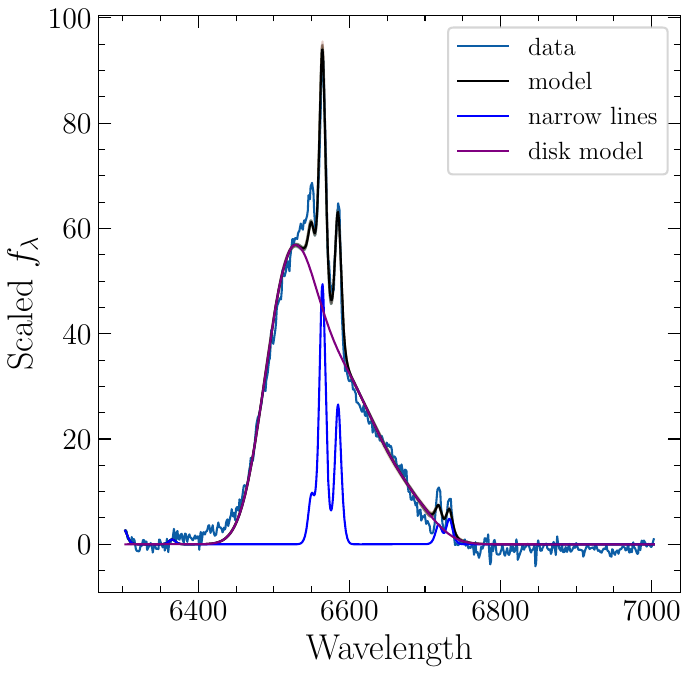}{0.24\textwidth}{h) BAT 862}}
\gridline{ \fig{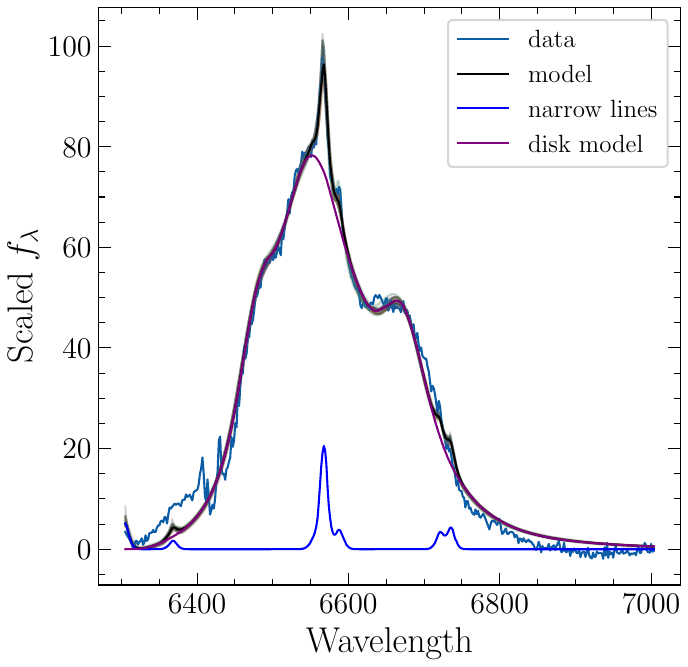}{0.24\textwidth}{i) BAT 907} \fig{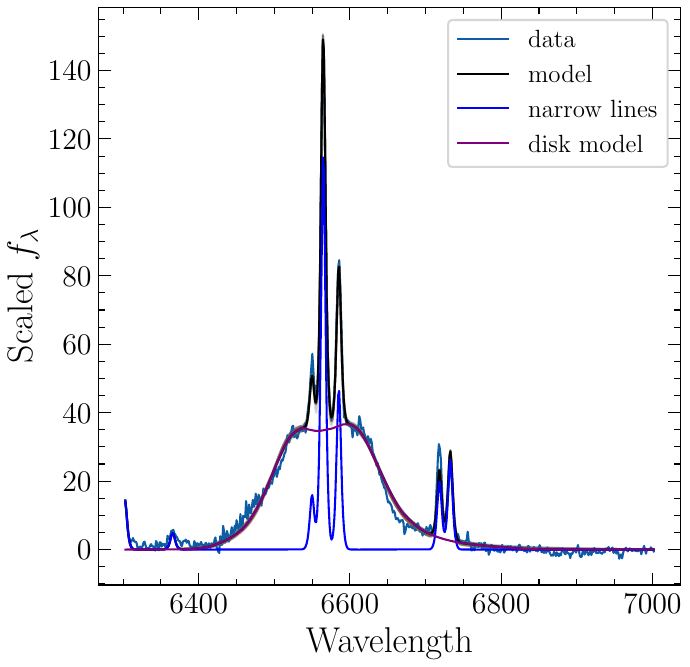}{0.24\textwidth}{j) BAT 1000}\fig{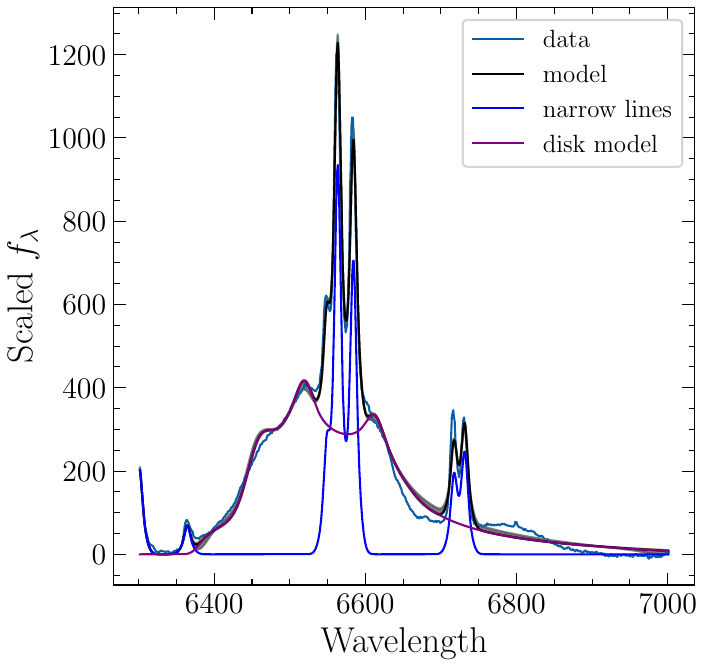}{0.24\textwidth}{k) BAT 1183}\fig{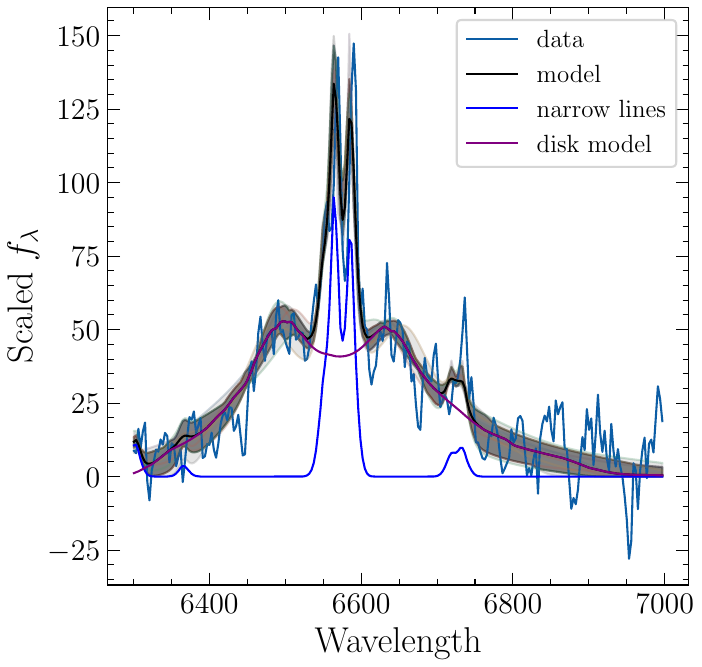}{0.24\textwidth}{l) BAT 1194}}
\caption{Examples of different disk profile morphologies for 12 DPEs identified in the parent sample for DPEs where an additional Gaussian broad line component was not required to produce a good fit. We show the data from the continuum-subtracted spectrum reported in Table \ref{table:diskparams} in blue, the best-fit narrow line component in dark blue, the disk model component in purple, and the summed model in black. The gray band around the total model shows the distribution of total model fluxes when taking 100 random samples of the MCMC walkers after convergence. It therefore reflects the range of possible models within the reported parameter uncertainties.}
\label{fig:profile_egs}
\end{figure*}

\begin{figure*}
\gridline{\fig{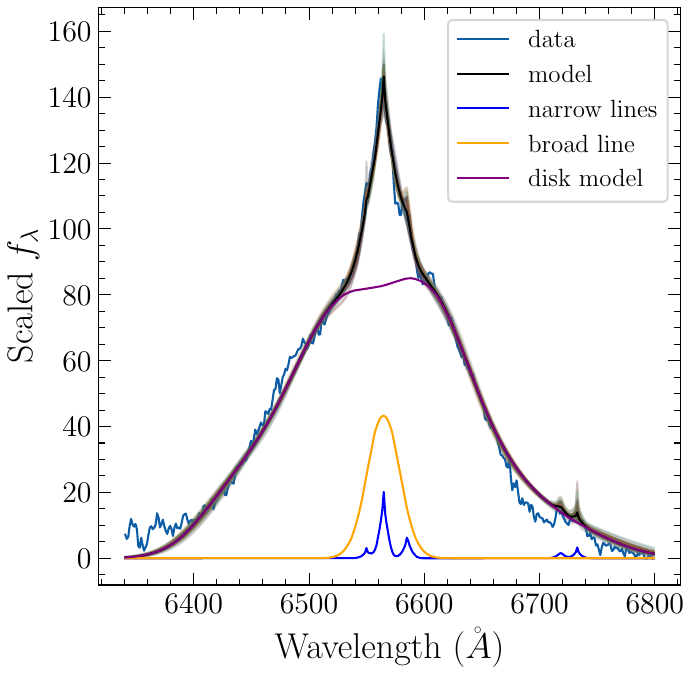}{0.24\textwidth}{a) BAT 121}\fig{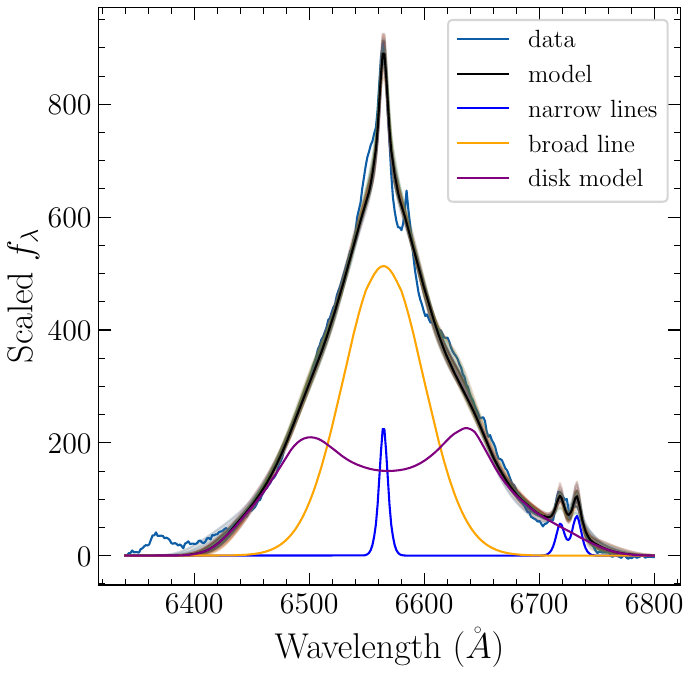}{0.24\textwidth}{b) BAT 162} 
\fig{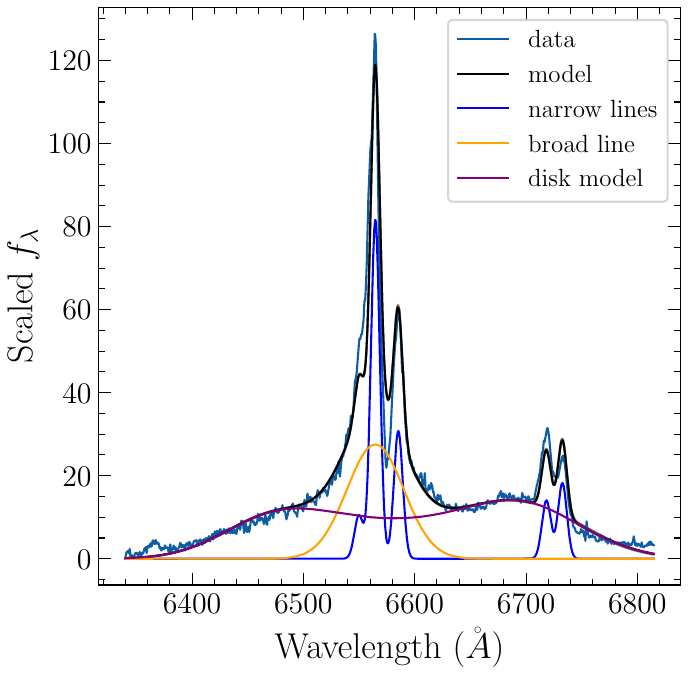}{0.24\textwidth}{c) BAT 315}\fig{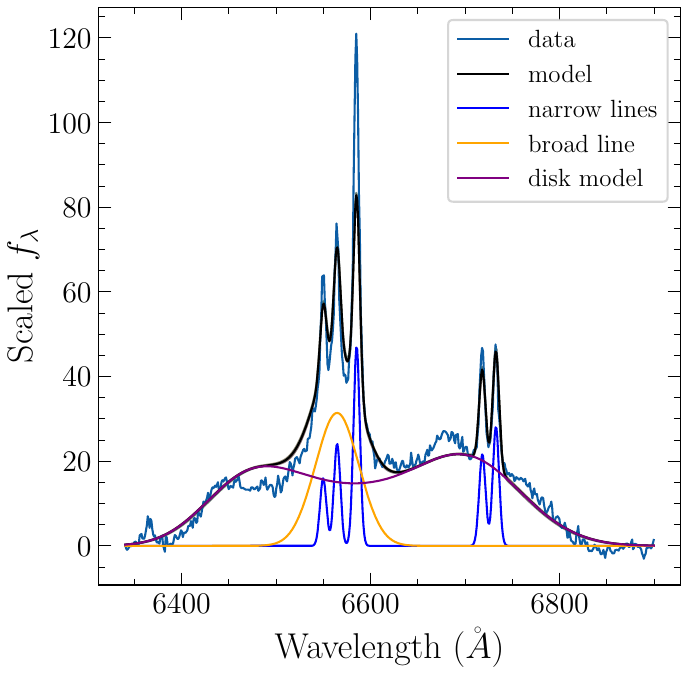}{0.24\textwidth}{d) BAT 395}}
\gridline{\fig{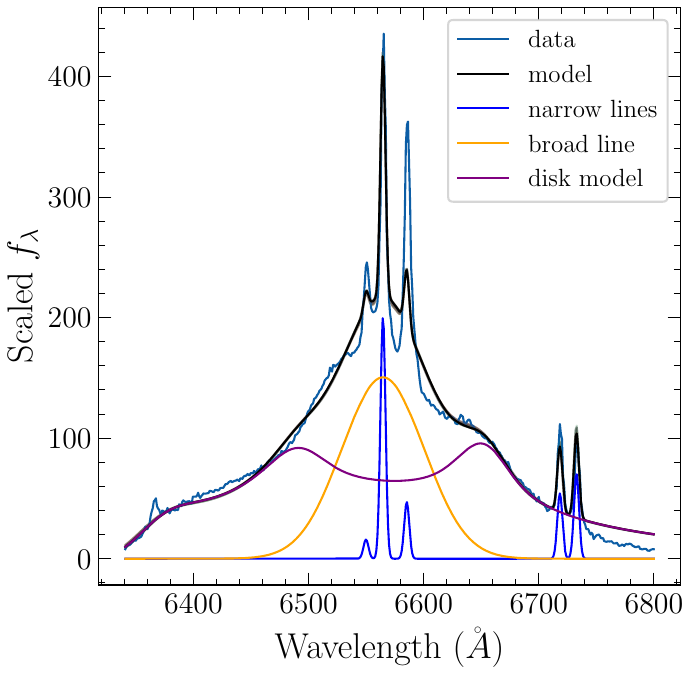}{0.24\textwidth}{e) BAT 420} 
\fig{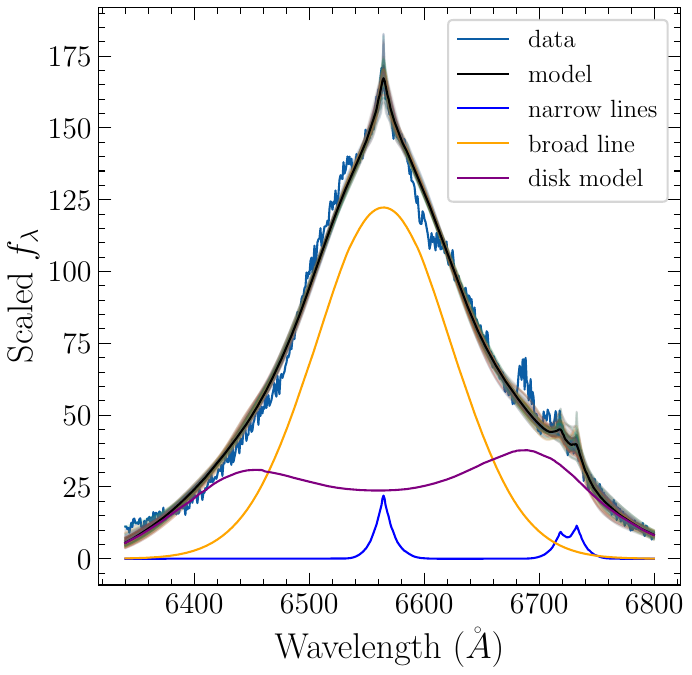}{0.24\textwidth}{f) BAT 431}\fig{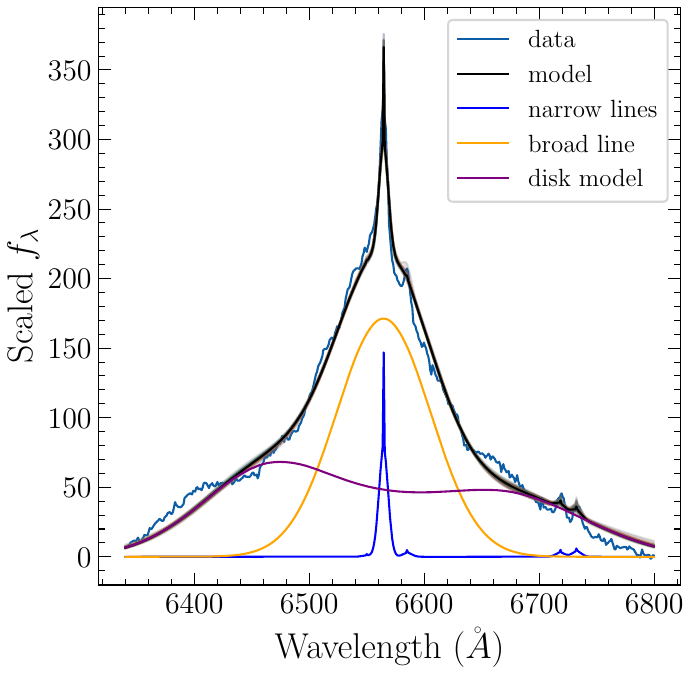}{0.24\textwidth}{g) BAT 473}\fig{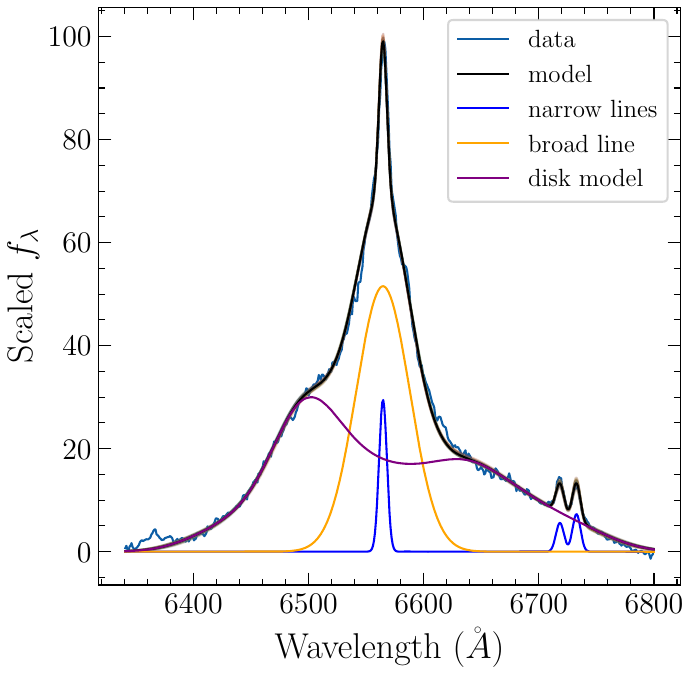}{0.24\textwidth}{h) BAT 537}}
\gridline{ \fig{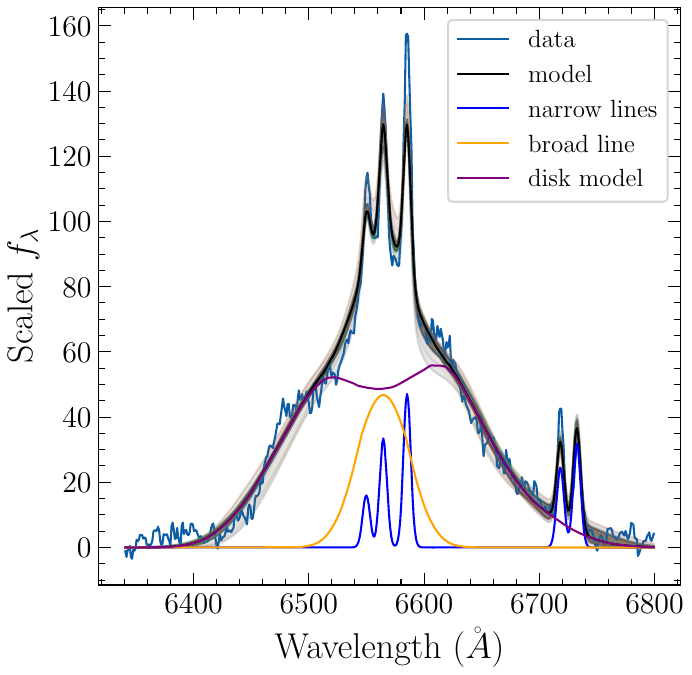}{0.24\textwidth}{i) BAT 633} \fig{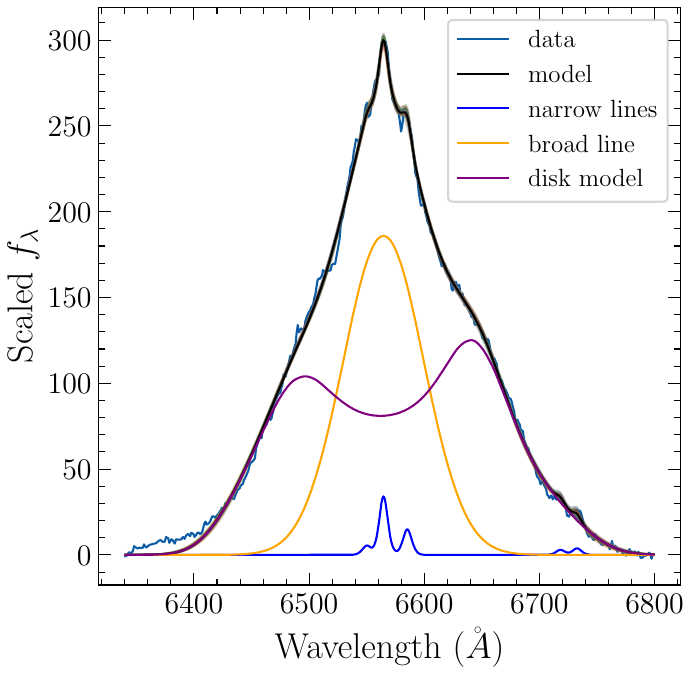}{0.24\textwidth}{j) BAT 748}\fig{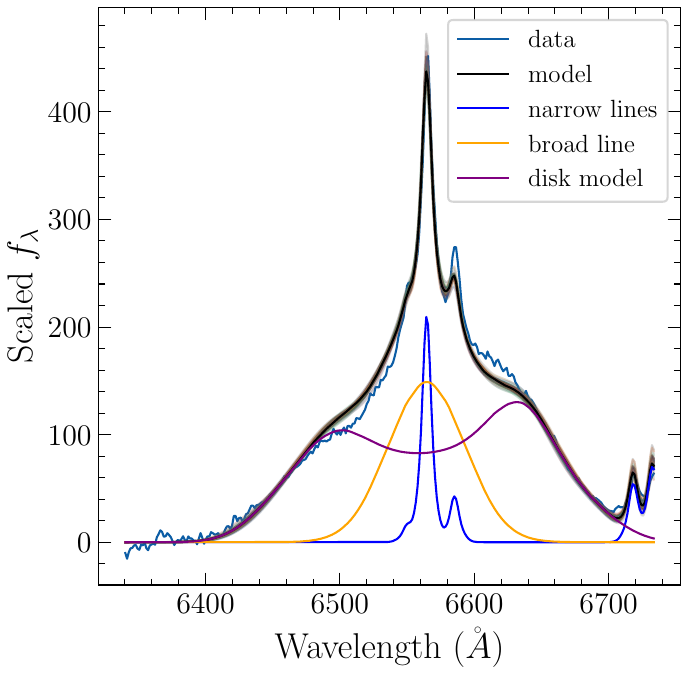}{0.24\textwidth}{k) BAT 1120}}
\caption{Disk profile morphologies for the 11 DPEs identified in the parent sample where an additional Gaussian broad line component was required to produce a good fit. We show the data from the continuum-subtracted spectrum reported in Table \ref{table:diskparams} in blue, the best-fit narrow line component in dark blue, the disk model component in purple, the additional Gaussian broad line in orange, and the summed model in black. The gray band around the total model shows the distribution of total model fluxes when taking 100 random samples of the MCMC walkers after convergence. It therefore reflects the range of possible models within the reported parameter uncertainties.}
\label{fig:profile_egs_addBL}
\end{figure*}

After continuum subtraction, we modeled the broad H$\alpha$ of each AGN with the circular accretion disk model from \cite{Chen1989}. We followed the procedure established for disk model-fitting of the ZTF AGN sample in \citet{Ward2024}, but we reiterate the procedure here for completeness. The disk model has the following free parameters: the inclination angle $i$ (deg) where 0\textdegree\ is face-on and 90\textdegree\ is edge-on, a local turbulent broadening parameter $\sigma$\ (km s$^{-1}$), the emissivity power law index $q$ where the emissivity varies with radius as $\epsilon \propto r^{-q}$, and the inner and outer dimensionless inner and outer radii of the disk $\xi_1$ and $\xi_2$ expressed relative to the gravitational radius, $GM_{BH}/c^2$. We enabled a single spiral arm with free parameters amplitude $A_s$ (expressed as a contrast ratio relative to the rest of the disk), orientation angle $\phi_0$ (deg), width $w$ (deg), and pitch angle $\psi$ (deg). This was required to describe the flux ratio of the red and blue shoulders being $>$1 in a fraction of spectra, which has been commonly observed in other disk emitters \citep[e.g.][]{StorchiBergmann2003}. We applied the following bounds on some parameters via a uniform prior: $\xi_1>50$, $w<80$, $0<\psi<60$, and $A_s<1$, based on typical parameters found for DPEs with detailed spiral arm modeling of multi-epoch spectra \citep[e.g.][]{Schimoia2012ShortNGC1097,Schimoia2017}. 

The disk model was fitted simultaneously with a model for the forbidden narrow emission lines overlapping the broad H$\alpha$ line. The [S\,{\sc ii}] $\lambda\lambda$6718, 6733, [N\,{\sc ii}] $\lambda\lambda$6550, 6585, and [O\,{\sc i}] $\lambda\lambda$6302, 6366 doublet flux ratios (where we report vacuum wavelengths) were fixed to theoretical values of 0.77, 0.34, and 3.0 respectively \citep{Osterbrock2006}. We note that the [S\,{\sc ii}] ratio can vary between 0.4 and 1.4, but we choose to fix it to reduce the number of free parameters, as it is usually close to the wing of the H$\alpha$ line, where it does not significantly affect the disk parameters. Each individual narrow line was described by two Gaussian components of the same central wavelength with three free parameters which were common for all narrow lines: the velocity dispersion of the first Gaussian component $\sigma_1$, the velocity dispersion of the second Gaussian component $\sigma_2$, and the flux ratio of the two components $f_1/f_2$. The amplitudes of the spectral lines are linear parameters, and so for computational expediency, we used a profile likelihood technique in which, for a given set of narrow line, broad line, and disk model parameters, we determined the amplitudes via least squares optimization. 

We first found a reasonable initial fit using the nonlinear least-squares optimization implemented in \textsc{Python} using the \textsc{scipy} package. We then explored the posteriors using \textsc{emcee} \citep{Foreman-Mackey2013ttemcee/ttHammer} with 60 walkers initialized at the best-fit values from the least-squares fit, distributed according to the 1$\sigma$ error found from the least-squares covariance matrix. For each spectrum, the emcee fitting was run for 2400 iterations after a burn-in time of 1800 iterations. The long burn-in phase was required for the chains to converge, given the large number of parameters. If no combination of parameters could account for the difference between the red and blue shoulder peak fluxes, we repeated the fitting prescription with greater freedom assigned to the spiral arm amplitude, increasing the amplitude from 1 to 10. While the spiral arm is well-motivated for a subset of DPEs and has been shown to be a physically motivated way to account for profile variability in a subset of DPEs \citep{Schimoia2017}, the spiral arm can introduce significant flexibility in the shape of the disk profile, so we did not enable it to have a large amplitude by default and only applied the additional freedom as necessary. If increasing the maximum allowed spiral amplitude still did not enable a reasonable fit to the data, and there was evidence for an additional single-peaked component at the rest wavelength, we repeated the fitting procedure while also allowing for an additional single-peaked broad-line component modeled as a Gaussian of adjustable width and flux centered on the H$\alpha$ rest velocity. 

Following the guidelines established in \citet{Ward2024} to separate double-peaked broad lines from typical broad lines with no evidence for shoulders or asymmetries, we classified objects as DPEs if they had an inclination angle $i>14$\textdegree, turbulent broadening $\sigma>600$ km s$^{-1}$, and inner radius $\xi_1<1200$. This procedure was established in \citet{Ward2024} as disk models that meet these requirements have broad line profiles where two separate shoulders are not blended. The distribution of these three parameters for the BASS sample, and the DPE cutoffs, are shown in Figure \ref{fig:disksplit}. This procedure led to 102 AGN classified as DPEs and the remaining 241 classified as `non-DPE' broad-line AGN, i.e. profiles with no evidence for two shoulders in the broad-line profile due to disk emission. We visually inspected the disk profile fits to the spectra, and reassigned 46 DPE candidates as likely AGN without disk profiles but instead with outflows causing asymmetries that produced disk model parameters consistent with a DPE. We reassigned 15 objects classified as non-DPEs to the DPE class, based on the presence of a dip in shoulders or a clear shoulder on the red or blue side of the profile. This resulted in the sample being split into 71 (21\%) DPEs and 272 (79\%) non-DPEs. 21 DPEs required a high amplitude spiral arm and 11 required an additional broad Gaussian component. Examples of BASS AGN classified as DPEs and their broad-line models are shown in Figure \ref{fig:profile_egs}. In addition, we show the models of 11 DPEs that required an additional Gaussian broad line in Figure \ref{fig:profile_egs_addBL}. We emphasize that this criteria may miss the identification of low inclination angle DPEs, where the shoulders are sufficiently close together such that the profile appears single-peaked. We may also miss DPEs where the combination of the radii and the level of turbulent broadening hide the presence of two shoulders. The sample of non-DPE broad-line AGN may therefore have some contamination from unidentifiable DPEs, and the 21\% DPE fraction should be considered a lower limit.

We note that in some cases (e.g. BAT 116, BAT 395) complex substructures in the broad line profile are not fully modeled by the simple disk model: further free parameters for disk substructures like spiral arms or disk winds would allow the models to better match the data, but we do not add these additional parameters to avoid overfitting. In the case of BAT 800, the dip on the red side of the H$\alpha+$ [NII] complex is caused by imperfect correction of the telluric ``A band''. As has been demonstrated in previous fits of Arp 102B and others, it is the width of the profile and the separation of the shoulders that are most constraining on inclinations and disk radii \citep{Chen1989, Strateva2003Double-peakedNuclei, Eracleous2009Double-peakedNuclei}, so we report the best-fit disk parameters for objects where these features of the profile shape were well-described by the disk model. For one object, BAT 744, no model was found that adequately described the blueshifted broad line, so we do not report its disk parameters. The DPE sample is summarized in Table \ref{table:ztfcands} and the non-DPE broad-line AGN sample (consisting of the remaining objects) is summarized in Appendix Table \ref{table:controlsample}. The best-fit disk parameters, details about their BASS spectrum, and an indication of whether an additional broad line was needed, are shown in Table \ref{table:diskparams}. We summarize the properties of the two samples in Figure \ref{fig:parentsample}, where we show the redshifts, the bolometric luminosities derived from intrinsic luminosity in the 14–150 keV range \citep{Ricci2017BATCatalog,Koss2022}, and the host galaxy morphology classifications from \citet{ParraTello2025}. The BH masses of the two samples are discussed in Section 5.


We note the distribution of Seyfert (Sy) 2s and intermediate Seyfert types as defined in \citet{Osterbrock2006} and reported for these objects in \citet{Oh2022}. Of the 71 objects classified as DPEs, 11 (15\%) have BASS spectral types as Sy1s, 17 (24\%) as Sy1.2s, 23 (32\%) as Sy1.5s, 6 (8\%) as Sy1.8s and 14 (20\%) as Sy1.9s. When looking at the total sample of 343 AGN, 11 of 47 Sy1s were DPEs (23\%), 17 of 84 Sy1.2s were DPEs (20\%), 23 of 137 Sy 1.5s were DPEs (17\%), 6 of 36 Sy1.8s were DPEs (17\%), and 14 of 52 Sy 1.9s were DPEs (27\%).

\begin{deluxetable*}{ccccccccccccc}
\tabletypesize{\scriptsize}
\tablecolumns{12}
\tablewidth{0pt}
\tablecaption{Properties of the 70 DPE candidates from the BASS broad-line AGN sample (full table available online) \label{table:ztfcands}}
\tablehead{
\colhead{ID} & \colhead{RA} & \colhead{Dec} & \colhead{z} &\colhead{Type} &\colhead{Galaxy} &\colhead{W2}& \colhead{W1-W2} &\colhead{ZTF }&\colhead{WISE }&\colhead{VLASS 3GHz}&\colhead{RACS 1GHz}\\[-0.3cm]
\colhead{} & \colhead{(hms)} & \colhead{(dms)} & \colhead{}& \colhead{}& \colhead{Morphology} &\colhead{(mag)}& \colhead{(mag)} &\colhead{variance}&\colhead{$\chi^2$/dof}&\colhead{(mJy)}&\colhead{(mJy)}}
\startdata
45&01:01:24.38&-03:08:40.20&$0.0696$&Sy1.5&disk-no-spiral&$11.6$&$0.56$&$0.03$&250.1&ND&D\\
61&01:13:50.09&-14:50:44.52&$0.0527$&Sy1.2&merger&$9.93$&$0.78$&$0.04$&740.94&$1.87\pm0.26$&$58.56\pm4.7$\\
107&02:07:02.21&$+$29:30:46.08&$0.1092$&Sy1.9&smooth&$10.9$&$1.0$&$0.03$&681.84&$22.49\pm0.8$&-\\
111&02:08:34.94&-17:39:34.92&$0.129$&Sy1.5&point-like&$10.48$&$0.89$&$0.1$&95.12&$153.74\pm0.5$&$44.22\pm3.65$\\
116&02:14:33.55&$+$00:46:00.12&$0.0263$&Sy1.2&disk-spiral&$9.73$&$0.44$&$0.13$&973.47&$4.0\pm0.36$&$6.4\pm1.2$\\
121&02:22:06.34&$+$52:21:05.76&$0.2$&Sy1.2&point-like&$11.2$&$1.05$&$0.12$&287.63&$2.8\pm0.59$&-\\
122&02:22:35.21&$+$25:08:14.64&$0.062$&Sy1.5&disk-spiral&$10.27$&$0.82$&$0.01$&456.02&$2.53\pm0.26$&$3.39\pm0.86$\\
136&02:38:19.73&-52:11:32.28&$0.0456$&Sy1&smooth&$10.08$&$0.77$&$-$&8543.39&-&$2.76\pm0.89$\\
147&02:44:57.70&$+$62:28:06.60&$0.0448$&Sy1&other/unc.&$7.35$&$1.09$&$0.04$&1004.02&$645.11\pm2.41$&-\\
162&03:00:04.32&-10:49:28.56&$0.0328$&Sy1.2&disk-no-spiral&$8.85$&$0.92$&$0.22$&2696.59&$8.84\pm0.33$&$26.52\pm2.62$\\
166&03:10:44.38&$+$32:39:29.16&$0.127$&Sy1&other/unc.&$10.0$&$1.0$&$-$&334.26&ND&-\\
187&03:35:22.58&$+$19:07:28.92&$0.189$&Sy1.8&point-like&$11.38$&$1.11$&$-$&455.38&$2.59\pm0.3$&$5.88\pm1.08$\\
232&04:40:47.71&$+$27:39:46.80&$0.0364$&Sy1.5&smooth&$11.14$&$0.63$&$0.14$&388.96&$2.57\pm0.26$&$6802.05\pm478.02$\\
236&04:43:46.80&$+$28:58:18.84&$0.0215$&Sy1.9&other/unc.&$9.86$&$0.74$&$0.11$&2204.5&$5.3\pm0.26$&D\\
270&05:19:49.73&-45:46:43.68&$0.035$&Sy1.2&smooth&$9.6$&$0.94$&$-$&1061.23&-&D\\
315&06:00:40.10&$+$00:06:18.36&$0.114$&Sy1.9&smooth&$9.13$&$1.12$&$0.23$&32.96&$35.71\pm0.31$&$2080.45\pm147.2$\\
367&07:23:53.04&-08:06:14.40&$0.146$&Sy1&point-like&$11.72$&$0.87$&$0.13$&146.07&ND&-\\
372&07:27:21.12&-24:06:32.40&$0.122$&Sy1.5&point-like&$10.16$&$1.12$&$-$&20.75&$11.73\pm0.36$&$68.26\pm5.48$\\
381&07:43:01.44&$+$80:26:26.16&$0.118$&Sy1.9&other/unc.&$11.34$&$1.09$&$0.06$&790.24&$3.02\pm0.38$&-\\
395&07:52:44.16&$+$45:56:57.48&$0.051$&Sy1.5&smooth&$11.06$&$0.33$&$0.1$&254.47&$39.39\pm0.62$&-\\
413&08:18:14.64&$+$01:22:27.12&$0.0894$&Sy1.9&disk-spiral&$10.54$&$0.98$&$0.21$&806.59&$9.1\pm0.27$&$3.48\pm1.21$\\
418&08:29:42.72&$+$41:54:36.72&$0.1264$&Sy1.5&smooth&$10.83$&$0.9$&$1.26$&2450.27&$4.33\pm0.35$&-\\
420&08:32:25.44&$+$37:07:36.12&$0.0922$&Sy1.2&smooth&$10.37$&$1.02$&$0.08$&2989.52&$5.92\pm0.24$&-\\
\enddata
\vspace{0.1cm}
\tablecomments{Properties of the 71 DPE candidates identified among the BASS broad-line AGN. Col. 1: Swift-BAT ID number from \citet{Koss2022} Cols. 2-3: RA and DEC of BAT counterpart from \citet{Koss2022}. Col 4: spectroscopic redshift from \citet{Koss2022}. Col. 5: Seyfert classification from \citet{Oh2022}. Col. 6: galaxy morphology classification from \citet{ParraTello2025}. Cols. 7 and 8: median W2 Vega magnitude and median W1-W2 color across the NEOWISE light curves. Col. 9: excess variance from the g-band ZTF light curves. Col. 10: $\chi^2\text{/dof}$ of the WISE W2 light curves. Col. 11: 2-4 GHz radio flux from VLASS for epoch 1 (2017-2018), with 2.5$"$ beam, where ND indicates radio non-detection and dash indicates that the source was not within the surveyed region. Col. 12: 1GHz radio flux from ASKAP-RACS, with 15$"$ resolution, where D indicates that the source is visible in publicly available RACS imaging but is not in fields covered by the epoch 1 catalog.}
\end{deluxetable*}

\begin{deluxetable*}{lllllllllllll}
\tabletypesize{\scriptsize}
\tablecolumns{14}
\tablewidth{0pt}
\tablecaption{Best-fit accretion disk parameters for the 71 objects classified as DPEs (full table available online). \label{table:diskparams}}
\tablehead{\colhead{ ID}&\colhead{Source}&\colhead{Date} &\colhead{$\xi_{1}$}&\colhead{$\xi_{2}$}&\colhead{$\sigma$ (km s$^{-1}$)}&\colhead{$i$ (deg)}&\colhead{$q$}&\colhead{$w$ (deg)}&\colhead{$A_s$}&\colhead{$\psi$ (deg)}&\colhead{$\phi_0$ (deg)}&\colhead{BL?}} 
\startdata
45&XSHO&2019-08-01&$200^{+10}_{-10}$&$1310^{+290}_{-510}$&$1410^{+70}_{-60}$&$13^{+1}_{-1}$&$2.4^{+0.1}_{-0.1}$&$81^{+12}_{-6}$&$0.9^{+0.1}_{-0.1}$&$56^{+21}_{-22}$&$294^{+26}_{-29}$&N\\
61&DBSP&2019-08-02&$120^{+10}_{-10}$&$3120^{+20}_{-20}$&$930^{+90}_{-70}$&$17^{+1}_{-1}$&$1.9^{+0.1}_{-0.1}$&$86^{+5}_{-3}$&$1.0^{+0.1}_{-0.1}$&$26^{+16}_{-14}$&$221^{+350}_{-48}$&N\\
107&ARCH&2018-08-24&$360^{+30}_{-40}$&$1280^{+370}_{-560}$&$1580^{+50}_{-70}$&$21^{+1}_{-1}$&$2.4^{+0.2}_{-0.1}$&$44^{+12}_{-16}$&$1.0^{+0.1}_{-0.1}$&$51^{+89}_{-28}$&$336^{+14}_{-212}$&N\\
111&DBSP&2019-08-23&$260^{+20}_{-20}$&$3860^{+90}_{-100}$&$1670^{+50}_{-50}$&$22^{+1}_{-1}$&$2.1^{+0.1}_{-0.1}$&$89^{+2}_{-1}$&$1.0^{+0.1}_{-0.1}$&$65^{+18}_{-16}$&$294^{+18}_{-24}$&N\\
116&XSHO&2017-12-05&$220^{+10}_{-10}$&$3990^{+10}_{-10}$&$850^{+20}_{-20}$&$20^{+1}_{-1}$&$2.2^{+0.1}_{-0.1}$&$90^{+1}_{-1}$&$1.0^{+0.1}_{-0.1}$&$40^{+8}_{-7}$&$224^{+14}_{-11}$&N\\
121&DBSP&2017-08-31&$400^{+50}_{-80}$&$9730^{+400}_{-200}$&$1250^{+100}_{-110}$&$36^{+2}_{-4}$&$14.4^{+1.8}_{-1.8}$&$0^{+1}_{-1}$&$14.8^{+4.7}_{-6.7}$&$2^{+1}_{-1}$&$1^{+1}_{-1}$&Y\\
122&DBSP&2016-10-02&$240^{+50}_{-50}$&$3580^{+590}_{-320}$&$900^{+180}_{-230}$&$25^{+2}_{-2}$&$1.8^{+0.3}_{-0.2}$&$66^{+26}_{-17}$&$0.8^{+0.3}_{-0.2}$&$41^{+83}_{-33}$&$342^{+72}_{-202}$&N\\
136&BNCH&2016-09-12&$220^{+40}_{-40}$&$1200^{+220}_{-560}$&$1590^{+100}_{-90}$&$13^{+1}_{-1}$&$1.9^{+0.7}_{-0.4}$&$83^{+12}_{-5}$&$0.9^{+0.1}_{-0.1}$&$61^{+22}_{-21}$&$288^{+30}_{-27}$&N\\
147&DBSP&2016-10-02&$110^{+30}_{-30}$&$2030^{+260}_{-330}$&$2740^{+180}_{-140}$&$27^{+1}_{-2}$&$1.5^{+0.1}_{-0.2}$&$45^{+20}_{-26}$&$0.8^{+0.3}_{-0.1}$&$52^{+24}_{-25}$&$240^{+23}_{-51}$&N\\
162&DBSP&2018-08-23&$130^{+10}_{-10}$&$1010^{+60}_{-30}$&$600^{+30}_{-20}$&$13^{+1}_{-1}$&$22.2^{+1.0}_{-0.9}$&$1^{+1}_{-1}$&$84.2^{+5.1}_{-4.2}$&$20^{+1}_{-1}$&$8^{+1}_{-1}$&Y\\
166&DBSP&2017-08-31&$340^{+40}_{-40}$&$2570^{+160}_{-200}$&$1440^{+60}_{-70}$&$20^{+1}_{-1}$&$8.2^{+0.8}_{-1.5}$&$-0^{+1}_{-1}$&$16.5^{+5.8}_{-6.6}$&$4^{+1}_{-1}$&$3^{+1}_{-1}$&N\\
187&LRIS&2019-12-24&$60^{+10}_{-20}$&$9750^{+510}_{-190}$&$1370^{+110}_{-160}$&$28^{+1}_{-1}$&$1.6^{+1.1}_{-1.8}$&$-0^{+1}_{-1}$&$15.0^{+0.7}_{-0.4}$&$3^{+1}_{-1}$&$2^{+1}_{-1}$&N\\
232&DBSP&2018-09-10&$170^{+10}_{-10}$&$1200^{+200}_{-240}$&$1170^{+80}_{-70}$&$13^{+1}_{-1}$&$2.4^{+0.1}_{-0.1}$&$81^{+11}_{-6}$&$0.9^{+0.1}_{-0.1}$&$59^{+22}_{-22}$&$291^{+26}_{-26}$&N\\
236&DBSP&2017-08-31&$680^{+360}_{-180}$&$860^{+140}_{-110}$&$1410^{+100}_{-80}$&$16^{+1}_{-1}$&$1.8^{+0.5}_{-0.5}$&$86^{+6}_{-3}$&$9.6^{+0.4}_{-0.3}$&$30^{+47}_{-50}$&$259^{+8}_{-7}$&N\\
270&BNCH&2016-09-11&$1130^{+470}_{-600}$&$4000^{+10}_{-10}$&$2630^{+320}_{-220}$&$57^{+13}_{-23}$&$1.8^{+0.6}_{-0.5}$&$45^{+31}_{-31}$&$0.5^{+0.4}_{-0.3}$&$7^{+35}_{-36}$&$108^{+578}_{-672}$&N\\
315&MAGE&2019-12-10&$270^{+110}_{-20}$&$590^{+20}_{-20}$&$2000^{+10}_{-80}$&$24^{+2}_{-1}$&$25.5^{+0.3}_{-0.4}$&$-0^{+1}_{-1}$&$40.7^{+0.4}_{-0.3}$&$5^{+1}_{-1}$&$7^{+1}_{-1}$&Y\\
367&DBSP&2018-03-28&$340^{+150}_{-230}$&$1900^{+10}_{-10}$&$1590^{+40}_{-60}$&$28^{+3}_{-3}$&$1.2^{+0.3}_{-0.7}$&$56^{+22}_{-24}$&$6.8^{+3.1}_{-2.3}$&$28^{+76}_{-40}$&$341^{+101}_{-224}$&N\\
372&XSHO&2019-03-19&$820^{+40}_{-50}$&$4000^{+10}_{-10}$&$1570^{+30}_{-10}$&$53^{+2}_{-2}$&$2.5^{+0.1}_{-0.1}$&$26^{+5}_{-9}$&$9.2^{+1.1}_{-0.6}$&$57^{+22}_{-21}$&$167^{+17}_{-30}$&N\\
381&DBSP&2016-01-12&$1040^{+850}_{-700}$&$3490^{+740}_{-370}$&$2190^{+700}_{-440}$&$26^{+5}_{-5}$&$1.5^{+0.5}_{-0.6}$&$47^{+33}_{-30}$&$0.5^{+0.4}_{-0.3}$&$41^{+68}_{-30}$&$287^{+79}_{-62}$&N\\
395&SDSS&2004-02-20&$240^{+60}_{-90}$&$3990^{+10}_{-10}$&$3000^{+10}_{-10}$&$19^{+2}_{-5}$&$2.5^{+0.1}_{-0.1}$&$78^{+21}_{-8}$&$4.8^{+1.7}_{-1.9}$&$14^{+1}_{-2}$&$175^{+37}_{-22}$&Y\\
413&BNCH&2017-10-02&$260^{+130}_{-380}$&$2950^{+10}_{-10}$&$2410^{+40}_{-50}$&$88^{+218}_{-181}$&$1.6^{+0.5}_{-0.5}$&$51^{+33}_{-28}$&$5.8^{+3.4}_{-3.3}$&$3^{+61}_{-63}$&$11^{+467}_{-468}$&N\\
418&SDSS&2001-12-23&$150^{+10}_{-10}$&$3900^{+50}_{-40}$&$2640^{+140}_{-150}$&$8^{+2}_{-1}$&$2.5^{+0.1}_{-0.1}$&$60^{+34}_{-22}$&$0.8^{+0.5}_{-0.2}$&$24^{+77}_{-46}$&$218^{+495}_{-580}$&N\\
420&SDSS&2002-02-07&$50^{+10}_{-10}$&$1590^{+40}_{-50}$&$720^{+460}_{-30}$&$27^{+1}_{-1}$&$35.4^{+0.9}_{-0.6}$&$-0^{+1}_{-2}$&$98.8^{+15.2}_{-1.9}$&$8^{+1}_{-1}$&$27^{+1}_{-1}$&Y\\
\enddata
\vspace{0.1cm}
\tablecomments{Best-fit disk parameters from modeling the broad H$\alpha$ line of the AGN with the circular accretion disk model from \cite{Chen1989} and with an additional spiral arm superposed \citep{Schimoia2017}. Cols. 1-11, respectively: Swift-BAT ID; instrument used for observation \citep{Oh2022}; date of observation; radius $\xi_{1}$ (gravitational radii); outer radius $\xi_{2}$ (gravitational radii); turbulent broadening $\sigma$ (km s$^{-1}$); inclination angle $i$ (deg); spiral arm width $w$ (deg); spiral arm amplitude expressed as contrast ratio $A_s$; spiral arm pitch angle $\psi$ (deg); spiral arm phase $\phi_0$ (deg); additional Gaussian broad line was fitted (Y/N).}
\end{deluxetable*}

We note that a subset of the objects in Table \ref{table:ztfcands} have been previously identified as DPEs. BAT 107 (3C 59), BAT 473 (3C 277), BAT 800 (3C 332), BAT 907 (PKS 1739+18C), BAT 994 (3C 390.3) were reported in \citet{Eracleous1994Doubled-peakedNuclei}. BAT 270 (Pictor A) and BAT 420 (CBS 74) were reported in \citet{Eracleous2003CompletionNuclei}. BAT 443 (J090436) and BAT 1183 (J230443) were reported as DPEs in \citet{Strateva2003Double-peakedNuclei}. For the 5 objects with circular disk model parameters reported from spectra taken in 1991-1998 in \citet{Eracleous1994Doubled-peakedNuclei} and \citet{Eracleous2003CompletionNuclei} (BAT 107, BAT 800, BAT 994, BAT 270 and BAT 420), inclination angles reported from the fits to BASS spectra are consistent with previous fits within 3$\sigma$ uncertainties, and have comparable estimates of the inner and outer radii. We note that the small differences observed in some parameters can arise naturally from small changes to the observed profile shape, and the slightly different modeling approach, where we allow a small amplitude spiral arm to improve the fit. 

While there are no significant changes in the peak separations for the 5 objects when comparing the 1991-1998 spectra to the BASS spectra taken three decades later, we note that BAT 800 and BAT 994 both show an increase in the flux of the broad-line profile compared to the narrow emission lines, and BAT 994 shows an increase in the blue-to-red shoulder flux ratio, consistent with year-to-decade evolution observed in samples of DPEs in \citet{Schimoia2017} and \citet{Ward2024}. BAT420 does not exhibit a major change in the disk profile except for adding a broad bump blueward of the blue shoulder in the disk profile.

\section{Variability properties of BASS DPEs and other broad-line AGN}

\begin{figure*}
\gridline{\fig{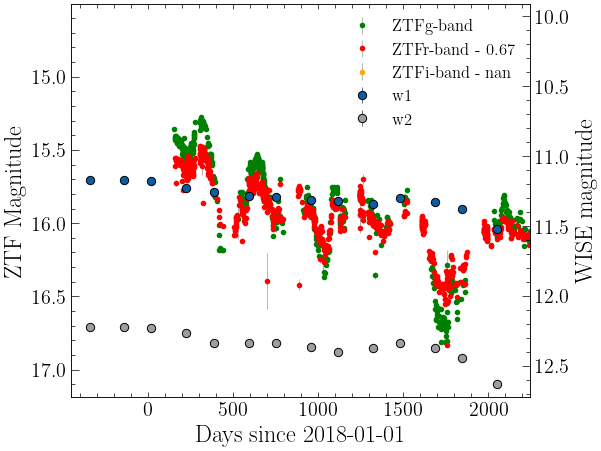}{0.48\textwidth}{a) BAT 121} \fig{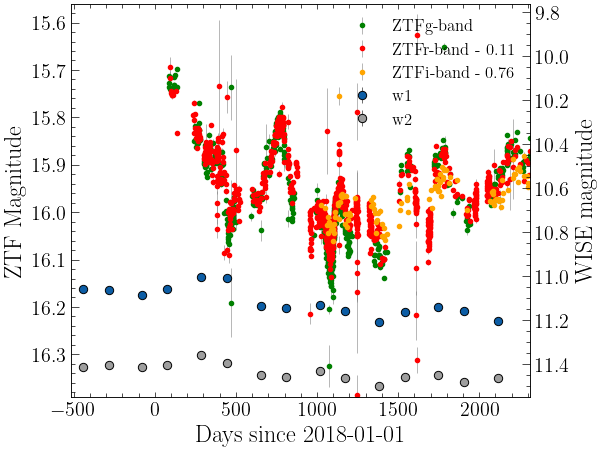}{0.48\textwidth}{c) BAT 395} } 
 \gridline{\fig{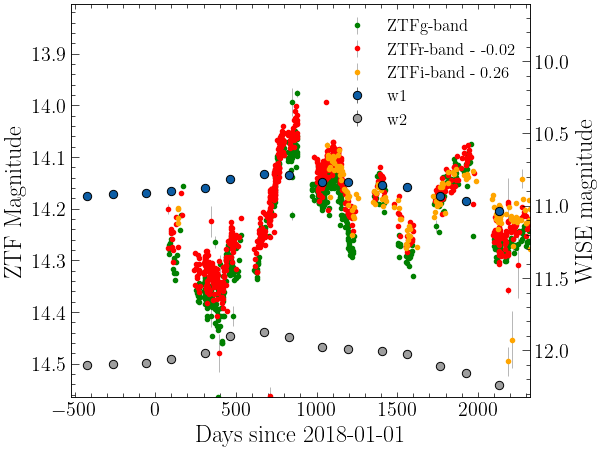}{0.48\textwidth}{d) BAT 429}\fig{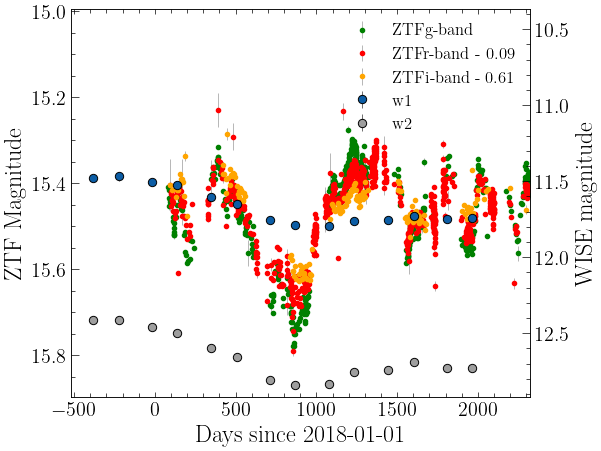}{0.48\textwidth}{e) BAT 770}}
\gridline{  \fig{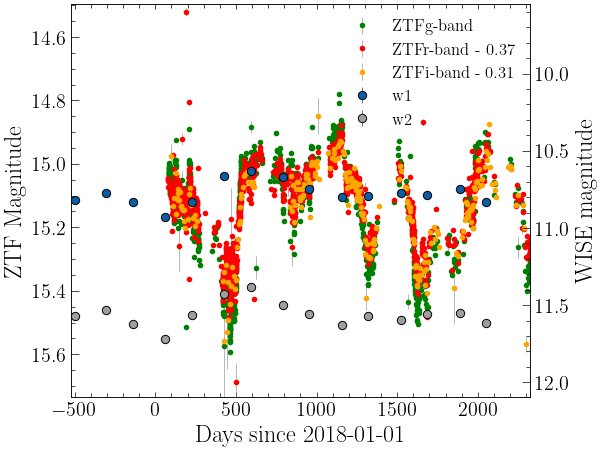}{0.48\textwidth}{f) BAT 862}\fig{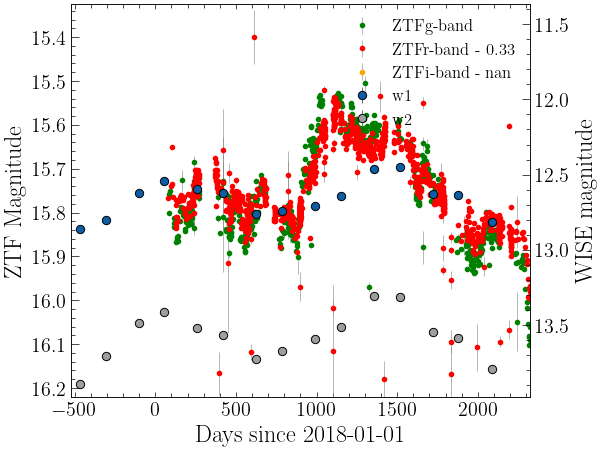}{0.48\textwidth}{b) BAT 1000}}
\caption{Optical (ZTF) and mid-IR (\textit{WISE}) light curves of six BASS DPEs. The left y-axes display the ZTF AB magnitudes while the right y-axes display the \textit{WISE} Vega magnitudes.}
\label{fig:lcs_notable}
\end{figure*}

Motivated by previous studies suggesting a) a high fraction of DPEs amongst optically variable AGN \citep{Ward2024} and b) different turnover frequencies between DPEs and other broad-line AGN \citep{Zhang2017PropertiesLines}, we make a comparison of the level of optical and mid-IR variability of the two populations. We constructed optical light curves of the BASS DPEs and non-DPE broad-line AGN sample from ZTF $g$ and $r$-band imaging using the ZTF forced photometry service, which produces point spread function (PSF) photometry from the ZTF difference images \citep{Masci2019TheArchive}. We extracted all available photometry from the ZTF public and partnership fields between 2018 January 1 and 2024 March 1. After removing poor quality images by requiring the \texttt{procstatus} flag be $=0$, we measured the baseline flux from  the reference images, applied zeropoints, and combined the baseline flux measured from the reference images and the single epoch fluxes to produce g- and r-band light curves of the two samples. 
We also produced mid-IR photometry available in W1 (3.4$\mu$m) and W2 (4.6$\mu$m) bands from the \textit{WISE} mission \citep{Mainzer2011NEOWISERESULTS, Mainzer2014InitialMission}. We obtained the NEOWISE light curves from IRSA \citep{NEOWISETeam2020NEOWISE-RTable}. NEOWISE observes each field with a $\sim6$ month cadence, taking multiple observations over a short $<2$ day period. We report the median and standard deviation of the observations taken upon each $\sim6$ monthly visit to the field. The mid-IR light curves of selected DPEs are also shown in Figure \ref{fig:lcs_notable}. 

In order to quantify the level of optical variability to facilitate comparison between DPEs and the non-DPE broad-line AGN sample, we calculated the excess variance of the ZTF g-band light curves, given by $\sigma^2_{NXS} = \frac{S^2 - \sigma^2_{n}}{\overline{f}^2}$, where $\overline{f}$ is the mean flux, $S^2$ is the total variance of the light curve, and $\sigma^2_{n}$ is the mean square photometric error associated with each measured flux \citep{Vagnetti2011}. We note that the galaxy light contribution is included in the reported light curves via addition of the reference image flux to the difference image fluxes, but our variability statistics capture the level of variability relative to the mean and will not be affected by the galaxy light contribution. For the WISE W2 light curves, we report the $\chi^2$/dof for a constant light curve at the median flux value, which was found to quantify AGN-like variability well in \citet{vanVelzen2021SeventeenStudies}, with $\chi^2/dof>10$ being a good indicator of AGN-like long-term variability. Table \ref{table:ztfcands} reports the optical and mid-IR variability metrics. The W2 $\chi^2/dof$ distributions and $g$-band excess variance are shown in Figures \ref{fig:compare}a and \ref{fig:compare}b. We applied a two-sample KS test to both parameters to determine the probability with which we can reject the null hypothesis that the non-DPE broad-line AGN and DPE parameters were drawn from the same distributions. For the optical excess variance and mid-IR $\chi^2/dof$ we obtain p-values of 0.61 and 0.70, indicating that we do not have evidence that they were drawn from different distributions (Table \ref{table:popdiff}). All AGN in both populations are variable in the optical and mid-IR.

\section{Comparing multi-wavelength properties of BASS DPEs and non-DPE broad-line AGN}

\begin{figure*}
\gridline{\fig{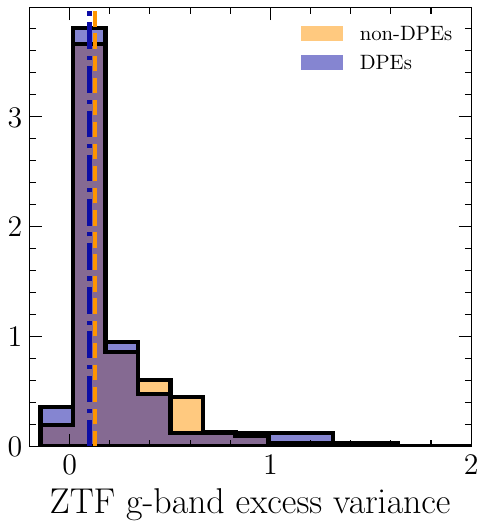}{0.3\textwidth}{a) g-band variability (ZTF)} \fig{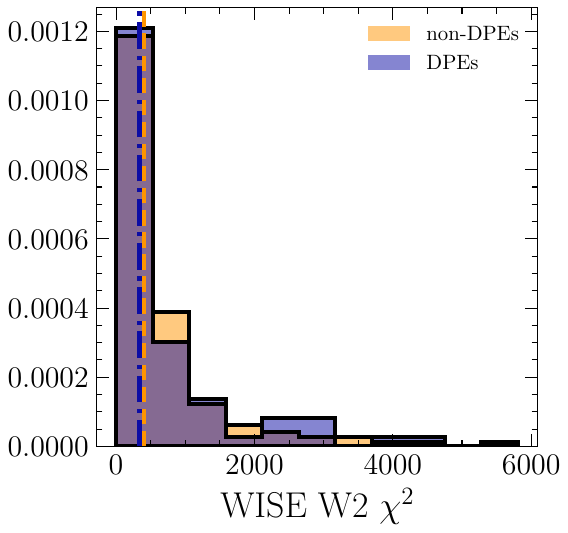}{0.33\textwidth}{b) W2 variability (WISE)} \fig{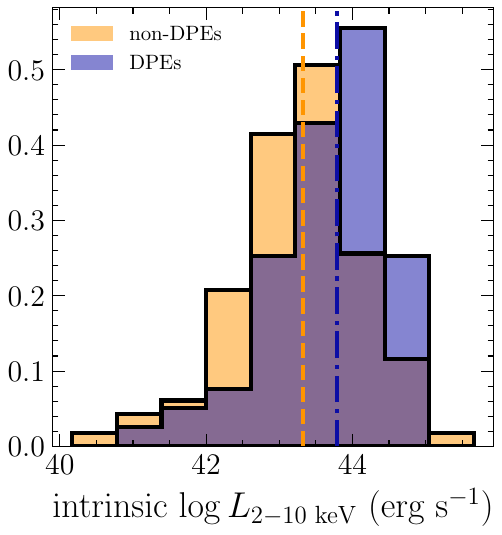}{0.3\textwidth}{c) Hard X-ray luminosity}  }
\gridline{\fig{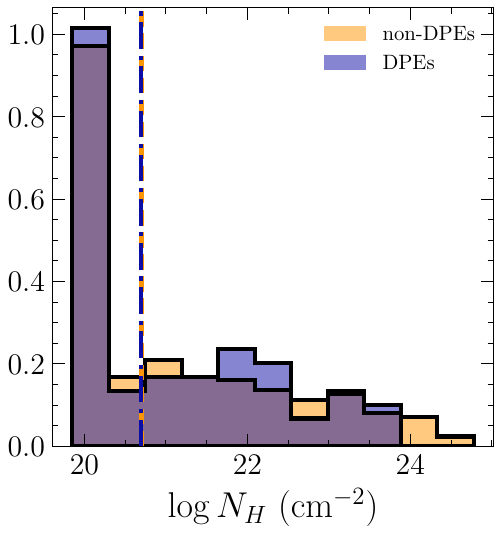}{0.3\textwidth}{d) Column density of obscuring material}
\fig{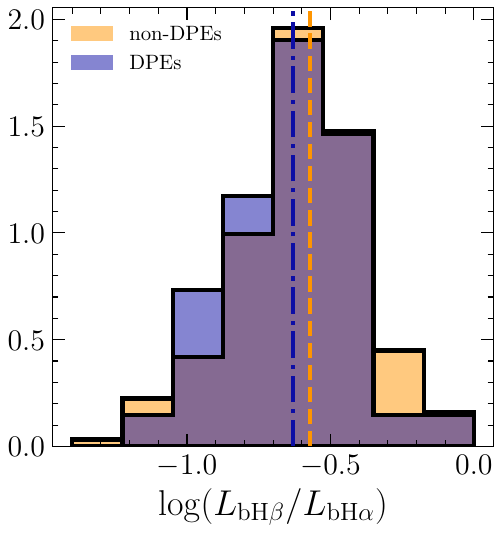}{0.3\textwidth}{e) Balmer decrement}\fig{KalphaEW.pdf}{0.33\textwidth}{f) Equivalent width of the Fe K$\alpha$ line}}
\gridline{ 
\fig{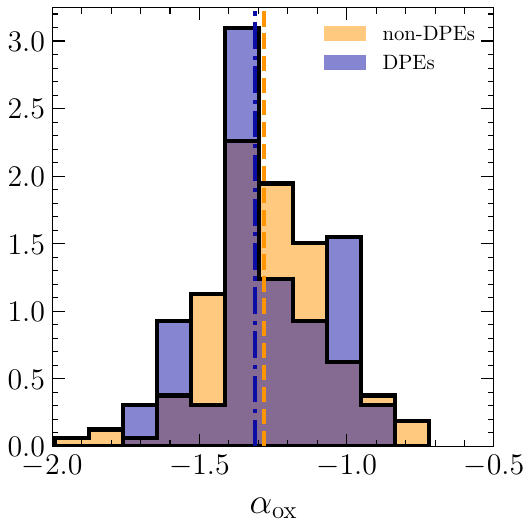}{0.31\textwidth}{g) Ratio of optical to X-ray luminosity}\fig{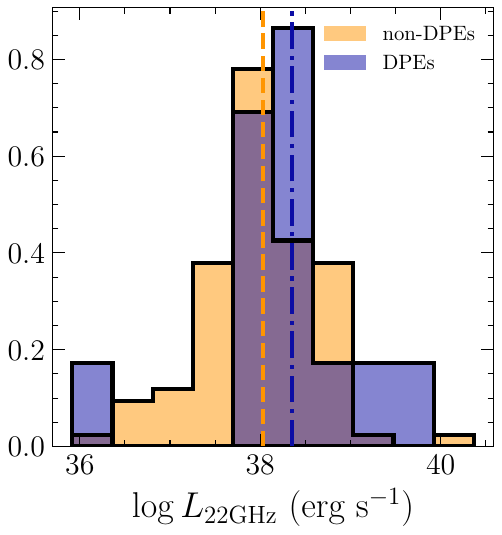}{0.31\textwidth}{h) 22GHz compact radio luminosity}\fig{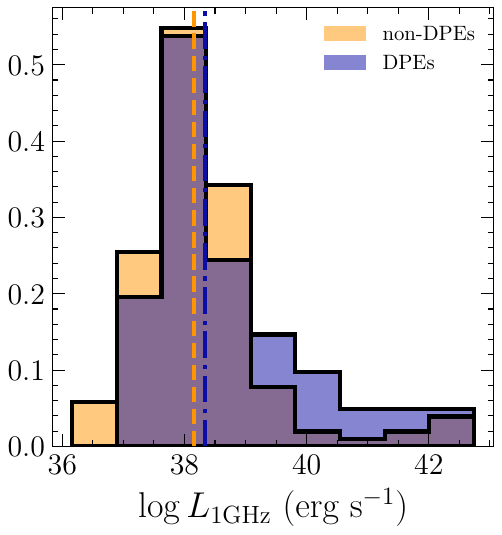}{0.31\textwidth}{i) 1GHz extended radio luminosity} 
} 
\caption{Normalized histograms comparing multi-wavelength properties of the BASS AGN after they are split into two classes: DPEs and non-DPE broad-line BASS AGN. For each histogram, we only show cases where that property was detected. The number of objects with detections going into each histogram are indicated in Table \ref{table:popdiff}. We indicate the median of each sample with a vertical purple dot-dashed line (DPEs) and an orange dashed line (non-DPEs).}
\label{fig:compare}
\end{figure*}

\begin{figure}
\gridline{\fig{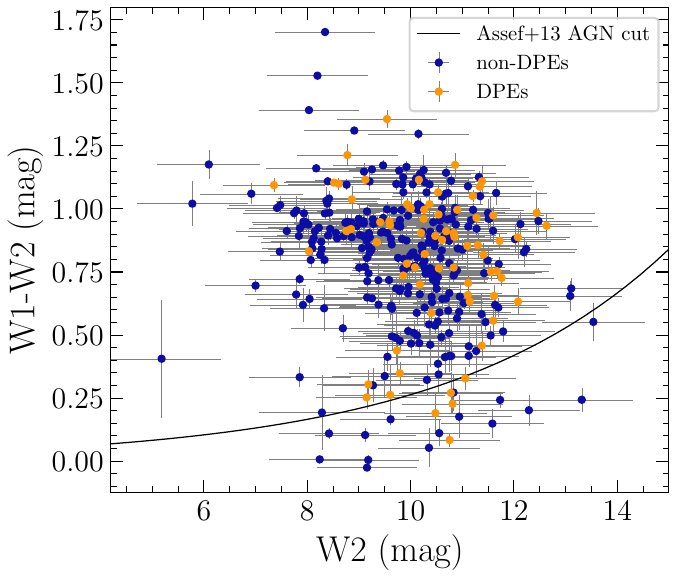}{0.45\textwidth}{}}
\caption{W1-W2 color vs W2 magnitude, and the \citet{Assef2013Mid-infraredField} cutoff for AGN. Magnitudes are Vega magnitudes. We show the BASS AGN split into two samples: DPEs and non-DPE BASS broad-line AGN.}
\label{fig:Wcolor}
\end{figure}

\begin{figure*}
\gridline{\fig{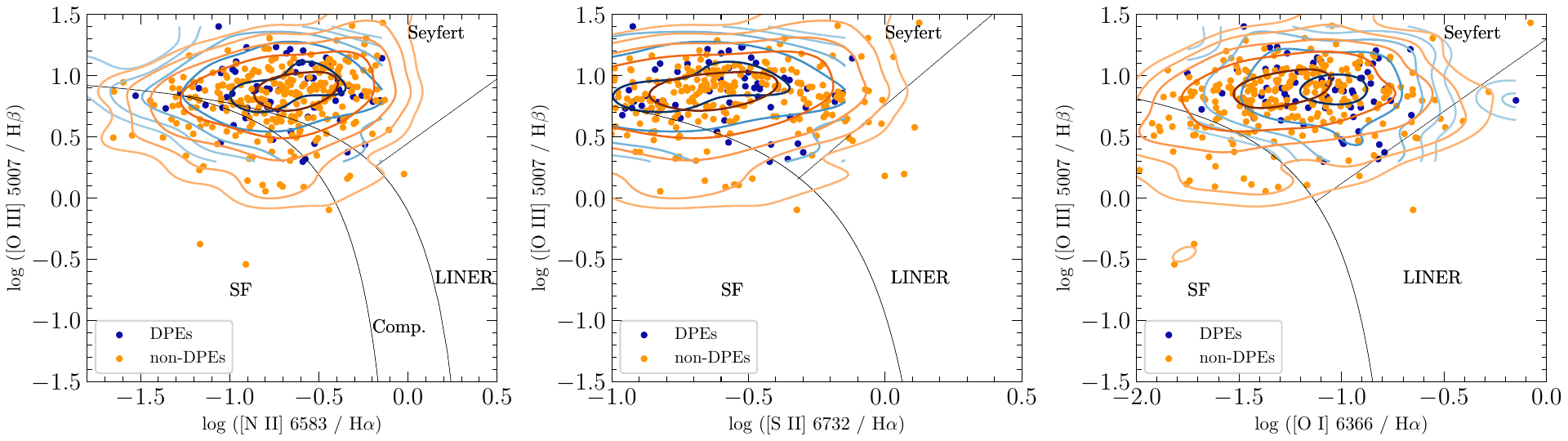}{0.99\textwidth}{}}
\caption{BPT diagrams showing the narrow emission line ratios derived in \citet{Oh2022} for BASS AGN split into two samples: DPEs and other broad-line AGN. The line diagnostics for different emission types \citep{Baldwin1981ClassificationObjects,Kauffmann2003TheNuclei,Kewley2001TheoreticalGalaxies,Kewley2006,Schawinski2007} are shown with black lines.}
\label{fig:bpt}
\end{figure*}

\begin{deluxetable*}{lllllll}
\tabletypesize{\scriptsize}
\tablecolumns{7}
\tablewidth{0pt}
\tablecaption{Comparison of DPEs and the non-DPE broad-line AGN sample for various multi-wavelength quantities  \label{table:popdiff}}
\tablehead{
\colhead{Quantity} & \colhead{\# DPEs} & \colhead{\# non-DPEs} & \colhead{DPE median} & \colhead{non-DPE median} & \colhead{KS statistic} &\colhead{p-value}}
\startdata
W1-W2 (Veg mag.) &70&280&0.88$\pm$0.27&0.85$\pm$0.26&0.08&0.827\\
WISE W2 excess variance (Vega mag.) &70&280&-0.00$\pm$0.03&-0.01$\pm$0.04&0.16&0.104\\
WISE W2 $\chi^2$ variance (Vega mag.) &70&280&338.97$\pm$1311.55&403.23$\pm$908.79&0.09&0.700\\
ZTF g-band excess variance (AB mag.) &58&207&0.10$\pm$0.26&0.13$\pm$0.45&0.11&0.612\\
nH (cm$^{-2}$)&66&278&20.70$\pm$1.17&20.70$\pm$1.33&0.07&0.921\\
$\alpha_{\text{ox}}$&28&138&-1.31$\pm$0.20&-1.28$\pm$0.20&0.11&0.928\\
SiVI Line Luminosity&3&28&39.83$\pm$0.69&40.22$\pm$0.74&0.38&0.709\\
FeK$\alpha$ EW (eV) &18&67&136.00$\pm$331.80&126.00$\pm$239.20&0.13&0.949\\
FeK$\alpha$ Energy (erg s$^{-1}$)&18&67&6.39$\pm$0.30&6.41$\pm$0.11&0.31&0.111\\
$\log (L_{\text{bH}\beta}/L_{\text{bH}\alpha})$&41&196&-0.63$\pm$0.29&-0.57$\pm$0.44&0.19&0.148\\
$\log M_{\text{BH}}$($M_{\odot}$) &18&85&8.18$\pm$0.45&7.80$\pm$0.58&0.39&0.015\\
$L_{3\text{GHz}}$ (erg s$^{-1}$)&52&176&29.53$\pm$1.16&29.43$\pm$1.08&0.14&0.344\\
$L_{22\text{GHz}}$(erg s$^{-1}$)&14&95&38.35$\pm$1.14&38.03$\pm$0.65&0.31&0.143\\
$L_{1\text{GHz}}$(erg s$^{-1}$)&28&140&38.33$\pm$1.29&38.16$\pm$1.08&0.27&0.056\\
$L_{2-10\text{ keV}}$(erg s$^{-1}$)&55&216&43.79$\pm$0.77&43.32$\pm$0.87&0.29&0.001\\
$\log L/L_{\text{Edd}}$ &17&81&-1.87$\pm$0.60&-1.56$\pm$0.58&0.32&0.075\\
OIb/H$\alpha$&71&285&0.08$\pm$0.09&0.05$\pm$3.26&0.31&0.000020\\
SIIb/H$\alpha$&71&285&0.25$\pm$0.17&0.22$\pm$329.63&0.11&0.471\\
NIIb/H$\alpha$&71&285&0.19$\pm$0.16&0.20$\pm$11737.19&0.08&0.778\\
OIIIb/H$\beta$&71&285&7.56$\pm$4.24&6.78$\pm$4715.37&0.13&0.240\\
\enddata
\vspace{0.1cm}
\tablecomments{Comparison of multi-wavelength properties of the BASS DPE sample and the sample of non-DPE broad-line AGN. Col. 1: quantity being compared and units for quoted medians. Cols. 2-3: number of DPEs and non-DPE broad-line AGN, respectively, with a measurement of that quantity. Col. 4: Median and standard deviation for the DPE sample. Col. 5: Median and standard deviation for the non-DPE sample. Col. 6: two-sample KS statistic. Col. 7: p-value for the probability with which we can reject the null hypothesis that the non-DPE broad-line AGN and DPE parameters are drawn from the same distributions.}
\end{deluxetable*}

To determine if the X-ray properties of the two samples differed, we compared the intrinsic X-ray luminosity $\log L_{\text{2-10keV}}$ (erg s$^{-1}$) and the column density of the neutral obscuring material derived from broad-band X-ray spectra $N_{\rm H}$, both reported in \citet{Ricci2017BATCatalog}. We find that the DPE population tends to be more X-ray luminous, with DPEs having a median and standard deviation  $\log L_{\text{2-10keV}}$ (erg s$^{-1}$) of $43.79\pm0.77$ and non-DPEs having a median and standard deviation  $\log L_{\text{2-10keV}}$ (erg s$^{-1}$) of $43.32\pm0.87$. While each population has a broad range of $\log L_{\text{2-10keV}}$ (erg s$^{-1}$) values, a KS-test to determine if the two samples have luminosities drawn from different distributions yielded a p-value of 0.001 (Figure \ref{fig:compare}c, Table \ref{table:popdiff}). Comparison of the $N_{\rm H}$ values, however, does not find any differences in the level of obscuring gas along the line of sight between the two populations (Figure \ref{fig:compare}d, Table \ref{table:popdiff}). We also compare the Balmer decrement  derived from the broad line luminosities obtained from multi-Gaussian fitting by \citet{Mejia2022} in order to search for differences in the physical conditions of the BLR dust, such as the density. DPEs have a slightly smaller $\log L_{\text{bH}\beta}/L_{\text{bH}\alpha}$ by 0.13 dex, but we cannot excluse that they are drawn from the same population with a KS test (Figure \ref{fig:compare}e, Table \ref{table:popdiff}). We note that uncertainties in the H$\alpha$ and H$\beta$ fluxes due to flux calibration issues reported in \citet{Mejia2022} may mean we have artificial biases in the reported Balmer decrements.

We include a comparison of the equivalent widths and energies of the Fe K$\alpha$ lines derived from X-ray spectroscopy analysis reported in \citet{Ricci2017BATCatalog}. The DPEs have Fe K$\alpha$ line energies and EWs which are typical of the BASS AGN sample as a whole (Table \ref{table:popdiff}, Figure \ref{fig:compare}f for EWs). We also show the distributions in the optical to X-ray flux ratio $\alpha_{\text{ox}}$ reported for this population in \citet{Gupta2024BASS.Nuclei}. We do not find any associations between DPEs and the X-ray/optical luminosity ratio (Figure \ref{fig:compare}g, Table \ref{table:popdiff}).

We compared the 22GHz radio luminosities from the galaxy nucleus measured from 1" resolution VLA imaging \citep{Magno2025}.  The distribution of $L_{\text{22GHz}}$ of the DPEs and non-DPE broad-line AGN with associated radio emission is shown in Figure \ref{fig:compare}h. The DPEs had a slightly higher median by 0.32, however, a two-sample KS test comparing the radio luminosities of the two populations has a p-value of 0.31 so we cannot rule out that they are from the same distribution for the 14 DPEs and 95 non-DPEs with 22 GHz imaging (Table \ref{table:popdiff}). 

We also compared the 3 GHz radio luminosities for all targets that were observed in the Karl G. Jansky Very Large Array Sky Survey at 2.5" resolution \citep[VLASS;][]{Lacy2020TheDesign}. We searched for crossmatches within $10^{\prime\prime}$ in Table 2 of the VLASS Epoch 1 Quick Look Catalogues which contains $\sim 700,000$ compact radio sources with $>1$ mJy/beam detections associated with mid-IR hosts from the un\textit{WISE} catalog \citep{Gordon2021ASurvey}. Out of the 65 DPEs in the VLASS survey area, 52 (80\%) had compact radio sources in VLASS. By comparison, for the broad-line AGN that were classified as non-DPEs, 166 out of 235 (71\%) were detected in VLASS. A two-sample KS test comparing the radio luminosities of the two populations has a p-value of 0.34, indicating no clear difference between the populations at 3 GHz (Table \ref{table:popdiff}). 

We also searched for radio emission in the Rapid ASKAP Continuum Survey (RACS), with first epoch observations covering the whole southern sky to +41 deg declination with the Australia Square Kilometre Array Pathfinder at a central wavelength of 887.5 MHz at $15^{\prime\prime}$ resolution \citep{Hale2021TheRelease}. We crossmatched our sample with a $10^{\prime\prime}$ radius to the first Stokes I Source Catalogue Data Release, which has an estimated 95\% point source completeness at an integrated flux density of $\sim3$ mJy. For any sources within the fields covered by the first data release that did not have a reported detection, we inspected imaging around the source coordinates with the CIRADA Image Cutout Web Service\footnote{http://cutouts.cirada.ca/}. For all observed sources without a reported detection in the catalog, an extended radio structure was clear in the CIRIDA imaging. Those sources are reported as detected at 1 GHz but we do not report a flux in Table \ref{table:ztfcands}. All DPEs except for BAT 1104 and all non-DPE broad-line AGN had radio sources in low-resolution RACS imaging when available. DPEs were more likely to have high 1 GHz luminosities in the $\log L_{22\text{GHz}}$ (erg s$^{-1}) = 39-43$ range in low-resolution 1 GHz imaging, although we once again do not find strong evidence that they were not drawn from the same distribution, with the KS-test p-value being 0.27 (Figure \ref{fig:compare}i, Table \ref{table:popdiff}). 

We also compare the mid-IR flux and colors of the two populations, using the median values and standard deviations of the W1 and W2 light curves described in Section 3. The distributions of the DPEs and the non-DPE broad-line AGN sample in the WISE W1-W2 vs W2 plot, with the classical AGN mid-IR color cut from \citet{Assef2013Mid-infraredField} overlaid, are shown in Figure \ref{fig:Wcolor}. The majority of both samples are classified as AGN on this plot, and no significant differences in the WISE color of the two populations are found (Table \ref{table:popdiff}). 

We compare the narrow emission line ratios reported in \citet{Oh2022} on three BPT diagrams \citep{Baldwin1981ClassificationObjects} in Figure \ref{fig:bpt}. We compare four line ratios for the two populations: the [O\,{\sc iii}] $\lambda$5007/H$\beta$ ratio, the [N\,{\sc ii}] $\lambda$6583 /H$\alpha$ ratio, the [S\,{\sc ii}] $\lambda$6732 /H$\alpha$ ratio, and the [O\,{\sc i}] $\lambda$6306 /H$\alpha$ ratio. The two-sample KS tests found evidence for differences between the two populations for the [O\,{\sc i}] $\lambda$6306/H$\alpha$ ratio with a p-value of $2\times 10^{-5}$. The median [O\,{\sc i}] $\lambda$6306/H$\alpha$ ratio is $\sim0.3$ dex greater for the DPE sample compared to the non-DPE sample. We discuss this further in Section 7. We note that for DPEs, the dip in the center of the H$\alpha$ and H$\beta$ broad line profiles can mean that the line ratios are inflated if the broad lines are modeled as Gaussians. This may account for the differences in the ionization ratios between the two populations derived in \citet{Oh2022}. We also compared the Si[VI] emission line detection fractions and luminosities measured in \citet{denBrok2022}, as this emission line is not contaminated by any broad double-peaked profiles. For the 23 BASS AGN with disk profile fits and Si[VI] line measurements, there were no differences between the DPEs and the non-DPE broad-line AGN sample, with 50\% of each sub-sample having emission lines, and with similar luminosity distributions between the two populations for those detected lines (Table \ref{table:popdiff}).

\section{BH masses of DPEs and non-DPE broad-line AGN}
\begin{figure}
\gridline{ \fig{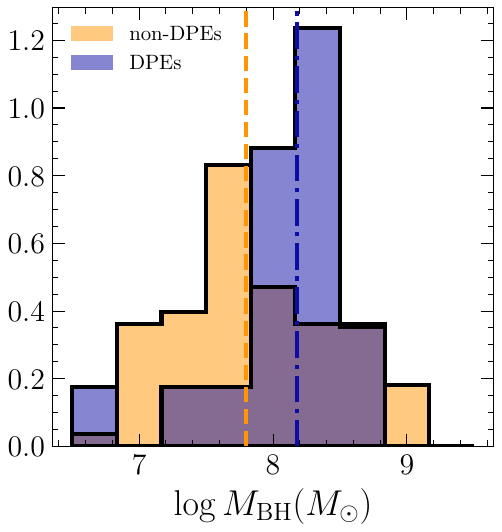}{0.25\textwidth}{}\fig{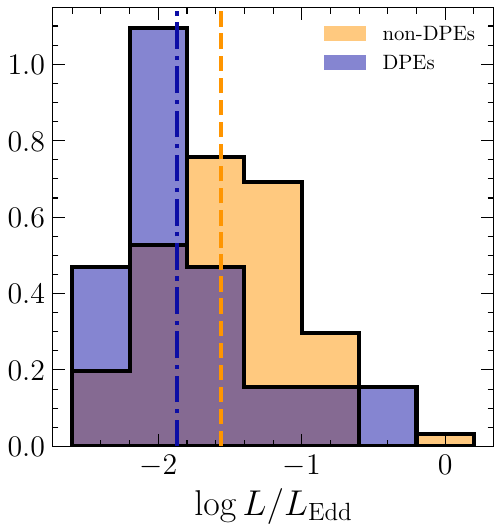}{0.25\textwidth}{}}
\caption{Comparison of masses (left) and derived Eddington ratios (right) when comparing DPEs and the non-DPE BASS broad line AGN. For both the masses and the Eddington ratios we used the BH mass extrapolated from the stellar velocity dispersion of the host galaxy. We indicate the median of each sample with a vertical purple dot-dashed line (DPEs) and an orange dashed line (non-DPEs).}
\label{fig:masscompare}
\end{figure}

\begin{figure}
\gridline{\fig{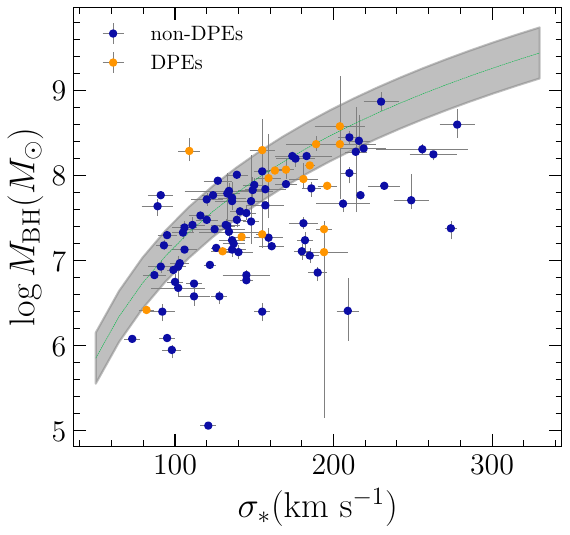}{0.45\textwidth}{}}
\caption{$M_{\text{BH}}-\sigma_*$ relation for BASS AGN with virial mass measurements \citep{Mejia2022} and stellar velocity dispersion measurements \citep{Caglar2023BASSAGNs}. The canonical relation from \citet{Kormendy2013CoevolutionGalaxies} with a scatter of 0.3 dex is shown in gray.}
\label{fig:Msig}
\end{figure}

By assuming virialized gas when deriving masses from the width of H$\alpha$ and H$\beta$ broad line profiles, virial masses will be overestimated for double-peaked emitters \citep{Eracleous2003CompletionNuclei}. We therefore must rely on extrapolations of BH mass from host stellar mass to identify if there are any intrinsic differences between the masses of DPEs and other broad-line AGN. We compared the black hole masses derived from the stellar velocity dispersion of the host galaxies reported in \citet{Caglar2023BASSAGNs} and applied the canonical \citet{Kormendy2013CoevolutionGalaxies} $M_{\text{BH}}-\sigma_*$ for the two samples. We find a significant preference for DPEs with higher mass MBHs, with DPEs having a median mass of $\log (M/M_{\odot})=8.2$ and the non-DPEs having a median mass of $\log (M/M_{\odot})=7.8$  (Figure \ref{fig:masscompare}, Table \ref{table:popdiff}). We also show the Eddington ratios derived from X-ray luminosity in \citet{Ricci2017BATCatalog} and BH mass inferred from stellar velocity dispersions in Figure \ref{fig:masscompare}. The median DPE $\log L/L_{\text{Edd}}$ is lower by 0.31. 


 In order to investigate how possible biases in virial mass measurements based on DPE broad line profiles may introduce scatter in the the $M_{\text{BH}}-\sigma_*$, we show where DPEs and other BASS broad-line AGN lie on the relation when using virial measurements from \citet{Mejia2022} and stellar velocity dispersion from \citet{Caglar2023BASSAGNs} (Figure \ref{fig:Msig}). Aside from the clear preference for larger BH masses, the DPEs do not stand out from the BASS AGN, which already show a preference for low virial masses, typically falling within or below the nominal 0.3 dex scatter observed about the   \citet{Kormendy2013CoevolutionGalaxies} relation.

\section{Host galaxy properties}
In order to compare the host galaxy properties of the DPEs to the other BASS AGN, we first generated a sub-sample of the main non-DPE broad-line AGN sample with comparable stellar mass and redshift to the DPEs. To do this, we used the host galaxy i-band magnitude reported in \citet{ParraTello2025} and undertook a sampling procedure to obtain a sub-sample of the non-DPE broad-line AGN with the same i-band host magnitude distribution as the DPE hosts. The sampling procedure was as follows: for each sample, we select a DPE at random, and then we randomly select one of the 15 non-DPE broad-line AGN with i-band magnitudes closest to that DPE's i-band magnitude. We repeated this procedure 300 times to produce a control sample of 300 AGN. We confirmed that this sample had comparable distributions of redshift, mass, and i-band magnitude as the DPE sample. 

We then compared the galaxy morphology classifications of the DPE sample and the i-band matched control sample, using the galaxy classifications reported in \citet{ParraTello2025}. We found that 33 of 71 DPEs (46\%) are classified as having `smooth' (elliptical) morphologies compared to the control sample's 48 of 164 ($31 \pm 2$\%), 6 of 71 DPEs (8\%) are in mergers compared to the control sample's 28 of 164 ($16\pm 1$\%), and 10 of 71 (14\%) DPEs are in `disk' galaxies - with or without a spiral - compared to the control sample's 43 of 164 ($29\pm2\%$). The population fractions and their 1$\sigma$ uncertainties reported for the control sample are derived from repeating the procedure to generate the control sample 20 times, and taking the mean and standard deviation of the population fraction each time. We discuss this further in Section 7.

\section{Discussion}
Our lower limit of a 21\% incidence rate of DPEs amongst BASS  X-ray--selected AGN of Seyfert types 1 to 1.9 is consistent with the detection rates previously found for a smaller sample of radio-loud broad-line AGN \citep{Eracleous2003CompletionNuclei}, and the 19\% detection rate of DPEs amongst populations of optically variable broad-line AGN \citep{Ward2021AGNsFacility}. It is substantially larger than the 3.6\% observed in a sample of SDSS quasars identified in optical spectroscopy \citep{Strateva2003Double-peakedNuclei}. 

Compared to the broad-line AGN sample, the low Eddington ratios and higher [O I]/H$\alpha$ line ratios are consistent with previous findings from smaller DPE populations \citep{Eracleous2003CompletionNuclei}. These findings have been explained by the expected SED of hot, vertically extended, optically thin, and radiatively inefficient accretion flow that can sustain radio jets and is more likely to appear at low accretion rates: the SED of such an accretion flow is relatively hard, having a power law shape across UV and soft X-ray bands with a peak in the far-IR, and lacking a UV bump \citep{DiMatteo1999,DiMatteo2000,Ball2001}. Such an ionizing continuum could produce the low ionization state of DPEs \citep{Nagao2002,Eracleous2003CompletionNuclei}. This supports a model that involves a transition in the BLR geometry to the observed disk-dominated state to produce such an SED.

We confirm that DPEs have a preference for elliptical galaxies (46\% compared to an $i$-band magnitude-matched control sample's 31\%), consistent with previous findings from \citet{Eracleous2003CompletionNuclei}. We confirm that DPEs tend to have host galaxies with larger stellar velocity dispersions. It has previously been suggested that the high black hole masses of DPEs may explain the high radio loudness fraction: more massive MBHs tend to be associated with smooth/elliptical galaxies  \citep{Eracleous2003CompletionNuclei, Strateva2003Double-peakedNuclei}. However, in this sample, we find that all DPEs except for 1 and all non-DPE broad-line BASS AGN that have low-resolution RACS imaging available are detected at 1 GHz, and that a greater fraction of the population have high 1 GHz integrated luminosities in the $\log L = 39-43$ erg s$^{-1}$ range than the non-DPEs. The DPEs in the sample had a slightly higher detection fraction in VLASS 3 GHz imaging (80\% instead of 71\% in the non-DPE sample), and did not have evidence for a difference in their 3 GHz luminosity distribution. When comparing compact 22 GHz compact emission in the 1$^{\prime\prime}$ nuclear region, DPEs tend to have slightly higher luminosities by $\sim 0.3$ dex. We note that the equivalent widths of the Fe K$\alpha$ lines are known to correlate with X-ray luminosity, radio-loudness, and Eddington ratio \citep{Page2004,Jiang2006}, and in such cases, there is no evidence for a difference between disk-emitters and the non-DPE broad-line AGN sample. 

\begin{figure}
\gridline{\fig{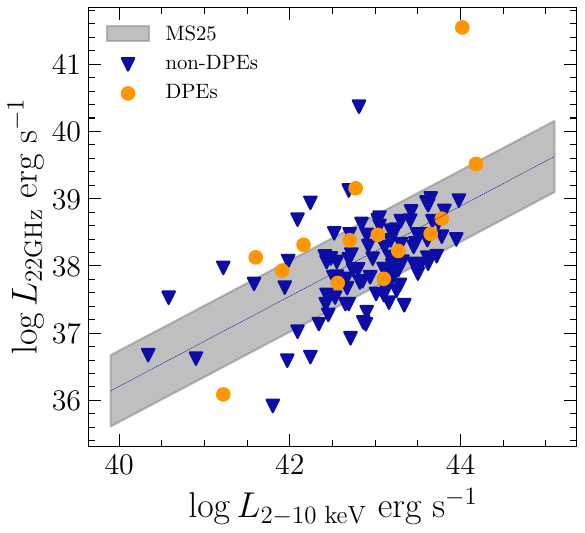}{0.46\textwidth}{}}
\caption{ Compact 22 GHz radio luminosity from \citet{Magno2025} vs the 2-10 keV X-ray luminosity reported in \citet{Ricci2017BATCatalog} for DPEs and non-DPE broad-line BASS AGN. In grey, we show the best-fit relation for BASS radio-detected BASS AGN from \citet{Magno2025}. No strong separation is seen between the two populations.}
\label{fig:XR}
\end{figure}

Given the higher [O I]/H$\alpha$ line ratios and possible association with a radiatively inefficient accretion flow that could sustain radio jets, it is surprising that we do not see a stronger preference for DPEs to have higher radio detection rates and luminosities in the various radio analyses. \citet{Terashima2003} previously found that the X-ray to radio luminosity ratio was able to produce a strong separation between Seyferts and low-luminosity AGN. In order to investigate whether DPEs show a preference for low $L_X$ to $L_R$ values, We plot the compact 22 GHz radio luminosity from \citet{Magno2025} against the 2--10\,keV X-ray luminosity of the two classes in Figure~\ref{fig:XR}. We overlay the best-fit relation found for the BASS AGN in gray, which was found by \citet{Magno2025} to be consistent with an extension of the relation found for coronally active stars \citep{Guedel1993} that extrapolates well to radio-quiet quasars \citep{Laor2008}. We note that both the DPEs and non-DPE BASS AGN are primarily radio-quiet, with the exception of one DPE and one non-DPE AGN with 22 GHz luminosities greater than 40 erg s$^{-1}$, which are classified as radio loud \citep{Magno2025}. We do not see separate clusters associated with the non-DPE broad-line AGN/DPE subclasses, and coronae are the primary drivers of compact 22 GHz emission in both populations.
According to the predictions of the disk-wind model of \citet{Elitzur2009OnNuclei, Elitzur2014EvolutionNuclei}, DPEs would be expected to preferentially appear in intermediate-type Seyferts over Seyfert 1s. This is because at lower accretion rates, more gas is expected to shift from high-altitude trajectories along the wind streamlines to low-altitude motion that follows the disk motion, causing a decrease in flux of single-peaked broad-line emission associated with winds and an increase in flux of a double-peaked disk emission line profile. Amongst our sample, 11 of 47 Sy1s were DPEs (23\%), 17 of 84 Sy1.2s were DPEs (20\%), 23 of 137 Sy 1.5s were DPEs (17\%), 6 of 36 Sy1.8s were DPEs (17\%), and 14 of 52 Sy 1.9s were DPEs (27\%). We therefore do not see strong correlations between DPE fraction and the broad-to-narrow line ratio in this systematically classified sample of AGN. 

\begin{figure}
\gridline{\fig{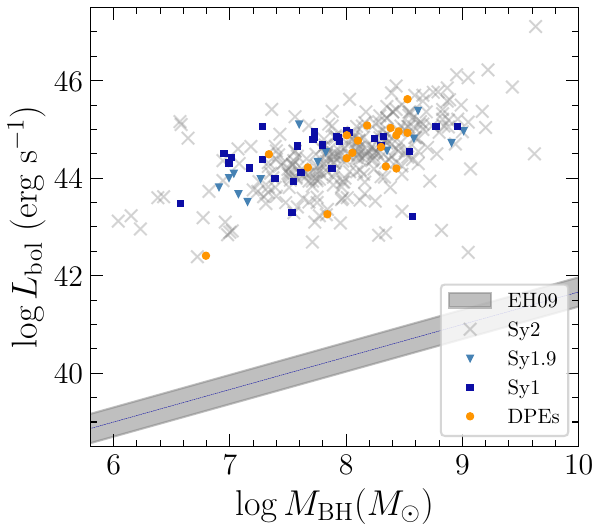}{0.49\textwidth}{}}
\caption{AGN bolometric luminosity derived from X-ray luminosity vs. BH mass estimated from stellar velocity dispersions of host galaxies for 4 sub-samples of the BASS AGN: DPEs, Sy2s and Sy1.9s that did not pass the DPE criteria, and the BASS Sy2 sample. The black line denotes the \citet{Elitzur2009OnNuclei} critical luminosity boundary expected to be the threshold below which the BLR disappears in the disk-wind model, while the grey shaded error region denotes a range of 0.3 dex based on typical scatter in bolometric luminosities derived from X-ray luminosities \citep{Ho2009b}.}
\label{fig:LbolM}
\end{figure}

In Figure \ref{fig:LbolM}, we show where the BASS AGN lie on the $L_{\text{bol}}$ vs $M_{\text{BH}}$ relation using the BH masses derived from stellar velocity dispersions and the bolometric luminosities derived from intrinsic luminosity in the 14–150 keV ranges \citep{Ricci2017BATCatalog,Koss2022}. We also show the boundary where the BLR emission is expected to disappear as the low radiative efficiency results in an advection-dominated state such that the BLR cannot be sustained \citep{Elitzur2009OnNuclei}. The BASS sample, including Sy2s shown in Figure \ref{fig:LbolM}, does not reach sufficiently low luminosities to enter this boundary region where \citet{Elitzur2009OnNuclei} and \citet{Elitzur2014EvolutionNuclei} observe only Sy2s and no broad-line AGN for the Palomar spectroscopic survey AGN sample. However, we note that there is no clear segregation between different Seyfert types or the DPE sub-sample based on accretion rate. In considering the disk-wind paradigm for disk-emitters, \citet{Elitzur2014EvolutionNuclei} noted that there were exceptions to their model, such as some high-Eddington ratio DPEs. It appears that within this X-ray-selected BASS sample, the disk-wind model cannot provide a clear explanation for the differences between DPEs and non-DPE broad-line AGN based on accretion rate alone.

We see no significant differences in the optical and mid-IR variability properties between the DPE and non-DPE broad-line AGN samples, in agreement with the variability-selected AGN sample from ZTF \citep{Ward2024}. This, in combination with the similar DPE rate for variable ZTF AGN, implies that variability amplitude does not assist in identification of DPEs -- a reflection that the two samples do not have large differences in optical luminosity, accretion rate, and black hole mass, which have been found to anti-correlate with variability amplitude \citep{Chanchaiworawit2024}. The lack of any significant differences in obscuration tracers such as mid-IR color and column density of obscuring material reflects that within the typical range of inclination angles $0<i<30$ for the two populations, we do not see much to distinguish the $i>14$\textdegree\ DPEs from the $i<14$\textdegree\ AGN, where the shoulders of a disk profile would not be distinguishable. 

\begin{figure}
\gridline{\fig{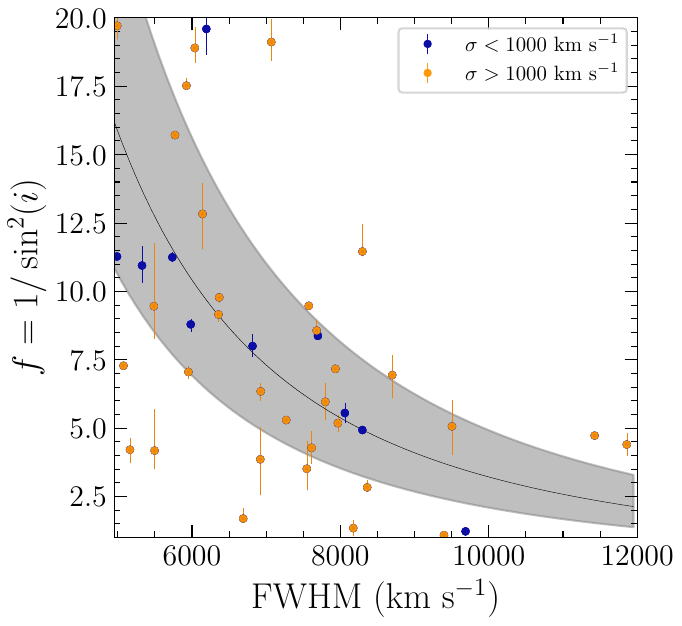}{0.45\textwidth}{}}
\caption{Relationship between the sine of the inclination angle and the FWHM of the broad line for the DPEs, split into populations with a small $\sigma<1000$ km s$^{-1}$ broadening parameter and a large $\sigma>1000$ km s$^{-1}$ broadening parameter. The black curve and grey shaded region denote the relation and dispersion found from simulations of disk profiles across inclination for DPEs with $\sigma=733 \pm 189$ reported in \citet{Storchi-Bergmann2017Double-PeakedNuclei}.}
\label{fig:FWHMinc}
\end{figure}

The presence of the disk profile can result in over-estimation of the virial BH mass measured from the broad line FWHM. Given the dependence of the profile width on disk parameters that are uncorrelated with mass, such as disk viewing angle, and parameters that may only be loosely correlated with BH mass, including the turbulent broadening parameter, it is difficult to obtain correct virial masses for DPEs. In Figure \ref{fig:FWHMinc} we show the relationship between the sine of the inclination angle and the FWHM of the broad line for the DPEs, split into populations with small ($\sigma<1000$ km s$^{-1}$) and large ($\sigma>1000$ km s$^{-1}$) broadening parameters. For comparison, we show the relation found from simulations of disk profiles across inclination with the average properties of the 5 DPEs in the sample of Type 1 Seyferts from the Palomar Spectroscopic Survey \citep{Filippenko1985,Ho1995}, which had $\sigma=733 \pm 189$, $\eta_1 = 1575 \pm 125$, and $\eta_2 = 3433 \pm 1211$ \citep{Storchi-Bergmann2017Double-PeakedNuclei}. For objects with FWHM within the range of 5000--12000 km s$^{-1}$ covered by the original paper, only those with intermediate inclinations angles and a low broadening parameter similar to the Palomar DPEs have a predictable relationship between FWHM and inclination angle. This demonstrates that both the inclination angle and the level of turbulent broadening --- and degeneracies between the two --- must be understood when identifying DPE candidates and correcting for the effect of the disk properties on the width of the broad line.

We note that two DPE candidates have been identified as changing-look AGN (CLAGN) or CLAGN candidates due to changes in their $H\alpha$ and H$\beta$ broad-line flux over time. BAT 1194 (IRAS 23226-3843) was classified as a changing-look AGN during a flaring event in 2019, where it showed double-peaked broad lines with a different shape compared to archival 1999 spectra \citep{Kollatschny2020}. BAT 862 (LEDA 1659236) was identified as a CLAGN candidate due to a decrease in broad line flux and a change in velocity structure observed in a 2020 spectrum compared to archival data in 2004 \citep{Temple2023}. The presence of 1--2 CLAGN in a sample of 71 DPEs is consistent with the 0.7–-6.2\% fraction observed for BASS AGN overall \citep{Temple2023}.

The differences that we observe between the DPE and non-DPE samples, including the high X-ray luminosities, low Eddington ratios, higher [O I]/H$\alpha$ line ratios, and earlier host type morphologies, point towards a physical difference in the BLR structure of the two populations. These differences may be even stronger if the unidentifiable disk-emitters at low inclination angles could be identified and removed from the non-DPE sample. The fact that 15\% of DPEs contained both a single and double-peaked contribution to the broad line indicates that some AGN are also in a transitional state between the two BLR geometries. To disentangle physical differences from viewing angle effects, further work on identifying face-on DPEs and DPEs with both a single and double-peaked component is needed. Given the degeneracies introduced by modeling both a double-peaked disk profile and a central Gaussian BLR, which fills the region between the peaks, it can be very difficult to determine if an AGN contains both components based on a single epoch spectrum. Multi-epoch spectroscopy may be an essential way to confirm disk emission in multi-component broad lines: in previous spectroscopic monitoring programs of AGN, analysis of the `root mean square' spectra, which often showed a double-peaked H$\beta$ profile implying the presence of variable, disk-like gas that may not be clear from a spectrum that appears single-peaked \citep{Denney2010ReverberationGalaxies, Schimoia2017,Storchi-Bergmann2017Double-PeakedNuclei}. Outflows also play a role in the quality of the disk models and DPE classifications: given that 55\% of Sy1.9s and 46\% of Sy1s in BASS DR1 had wings in the O[III] narrow lines indicating the presence of outflows \citep{Rojas2020}, an outflowing component may improve modeling of 50\% of the DPEs. 

Aside from identifying the variable disk components in `rms' spectra, analysis of the double-peaked profiles across UV/optical lines may be beneficial for understanding the different BLR components. Double-peaked near-IR lines also prove to be a useful probe: the recent detection of O I $\lambda$11297, Pa$\alpha$, and tentative He I $\lambda$10830 double-peaked broad lines with an additional Gaussian broad component in local Seyfert 1 galaxy III Zw 002 demonstrated this \citep{2023Santos}. Most recently, double-peaked profiles of the NIR Ca II triplet and O I $\lambda$8446 were identified in NGC 4593 \citep{Ochmann2025}. UV and near-IR spectroscopy of the BASS DPEs would enable more comprehensive disk modeling across multiple double-peaked line profiles.

Another opportunity to map the structure of the BLR region in DPEs is possible using the new capabilities of Resolve X-ray calorimeter \citep{Ishisaki2022} on the new X-ray Imaging and Spectroscopy Mission \citep[XRISM;][]{Tashiro2020}, which provides a resolution of $\Delta E \approx 4.5$ eV across the 1.6–17.4 keV pass band. XRISM/Resolve spectroscopy of the Fe K$\alpha$ complex of broad line AGN NGC 4151 showed that the line width was comparable with the broad H$\beta$ line \citep{Miller2024}. Modeling of the Fe K$\alpha$ line shape enabled separate detection of emission from the broad line region at $\xi \sim 3\times10^3 $, the inner edge of the dusty torus at $\xi \sim10^4$ , and warps at $\xi \sim 100$ \citep{Miller2024}. XRISM/Resolve observations of X-ray bright DPEs like BAT 1183 (Mrk 926) could search for a double-peaked broad component to the Fe K$\alpha$ line to aid interpretation of the BLR geometry implied by modeling of the Balmer lines.

\section{Conclusions}
In this paper, we have undertaken a population study of a hard X-ray selected and flux-limited sample of 343 Swift-BAT AGN from the BASS survey with optical spectroscopy indicating the presence of broad H$\alpha$ lines, providing a large sample of AGN that is unbiased to inclination angle or other selection effects introduced by optical spectroscopy. We fit the broad H$\alpha$ profiles with double-peaked disk models to systematically classify DPEs and AGN without double-peaked profiles, and report the disk properties of 71 DPEs (Table \ref{table:diskparams}), leading to the following findings.

Firstly, we find that $>$21\% of BASS AGN with broad H$\alpha$ emission have double-peaked disk profiles, similar to the rate found for variability-selected AGN \citep{Ward2024} and radio-loud broad-line AGN \citep{Eracleous2003CompletionNuclei}. As disk profiles introduce biases to virial mass measurements, this indicates that the presence of DPEs could introduce biases in the $M_\text{BH}-\sigma_*$ relation if they are unidentified in MBH populations. As the FWHM of the broad-line profile depends significantly on both inclination angle and the level of turbulent broadening in the disk (Figure \ref{fig:FWHMinc}), it is challenging to correct for the level of broadening introduced by the disk profile and caution must be taken when DPEs contaminate large samples used for $M_\text{BH}-\sigma_*$ studies.

Secondly, we confirm with this large, systematically classified sample that DPEs have intrinsically higher masses by $\sim 0.4$ dex and lower Eddington ratios by $\sim 0.3$ dex than non-DPE broad-line AGN when using BH masses estimated from stellar dispersion velocities, where the presence of the disk profile will not inhibit accurate mass measurement. DPEs also have a preference for elliptical hosts, higher X-ray luminosities -- with a median value that is 0.47 dex greater than non-DPEs -- and higher [O\,{\sc i}] $\lambda$6306/H$\alpha$ flux ratios when compared to the non-DPE broad-line AGN sample. The DPEs also had sub-arcsecond 22 GHz radio luminosities that were $\sim 0.3$ dex higher than the non-DPEs, and extended 1 GHz radio luminosities that were more likely to fall in the high $L_{1\text{GHz}} = 39-42$ (erg s$^{-1}$) range, but larger samples are required to find stronger evidence that the radio luminosities of the two populations are from different distributions. Despite the differences in masses, accretion rates, and X-ray and radio luminosities, DPEs are not significantly segregated from non-DPE broad-line AGN in the $L_{\text{bol}}$ vs $M_{\text{BH}}$ relation and do not show a preference for intermediate Seyfert types over Seyfert 1s, suggesting that accretion rate changes alone in the context of the disk-wind model may not be able to account for the transition to a disk profile. This finding for hard X-ray selected AGN contrasts with previous findings for spectroscopically-selected AGN samples \citep{Elitzur2014EvolutionNuclei}.

Finally, in this sample, we do not find differences across a wide range of multi-wavelength properties when comparing DPEs to non-DPE broad-line AGN, including optical and mid-IR variability levels, WISE colors, Fe K$\alpha$ equivalent width, the Balmer decrement, $\alpha_{\text{ox}}$, the column density of neutral obscuring material $N_H$, and the rate of changing-look events. The challenge of identifying DPEs at low inclination angles may be inhibiting clean separation of DPEs and non-DPE broad-line AGN. Further work is required to disentangle physical differences and viewing angle selection effects if we are to understand the different accretion geometries and observational properties of these systems.

\section{Acknowledgements}
We would like to thank the anonymous referee for their helpful comments. We acknowledge support from NASA through ADAP award 80NSSC22K1126 (MK), ANID Fondecyt Regular grants 1230345 (CR),1241005, and 1250821 (FEB, ET), ANID BASAL project FB210003 (CR, FEB, ET), Millennium Science Initiative - AIM23-0001 and ICN12\_009 (FEB), and the China-Chile joint research fund (CR). BT acknowledges support from the European Research Council (ERC) under the European Union's Horizon 2020 research and innovation program (grant agreement number 950533). IMC acknowledges support from ANID programme FONDECYT Postdoctorado 3230653. KO acknowledges support from the Korea Astronomy and Space Science Institute under the R\&D program (Project No. 2025-1-831-01), supervised by the Korea AeroSpace Administration, and the National Research Foundation of Korea (NRF) grant funded by the Korea government (MSIT) (RS-2025-00553982). AJ acknowledges support from the FONDECYT Postdoctoral fellowship 3230303.

This research has made use of the CIRADA cutout service at URL cutouts.cirada.ca, operated by the Canadian Initiative for Radio Astronomy Data Analysis (CIRADA). CIRADA is funded by a grant from the Canada Foundation for Innovation 2017 Innovation Fund (Project 35999), as well as by the Provinces of Ontario, British Columbia, Alberta, Manitoba and Quebec, in collaboration with the National Research Council of Canada, the US National Radio Astronomy Observatory and Australia’s Commonwealth Scientific and Industrial Research Organisation.

This publication also makes use of data products from NEOWISE, which is a project of the Jet Propulsion Laboratory/California Institute of Technology, funded by the Planetary Science Division of the National Aeronautics and Space Administration.

This research has made use of the NASA/IPAC Infrared Science Archive, which is funded by the National Aeronautics and Space Administration and operated by the California Institute of Technology.

\section{Appendix}

\begin{deluxetable*}{ccccccccccccc}
\tabletypesize{\scriptsize}
\tablecolumns{12}
\tablewidth{0pt}
\tablecaption{Properties of the 272 non-DPE broad-line AGN from the BASS broad-line AGN sample (full table available online) \label{table:controlsample}}
\tablehead{
\colhead{Swift-} & \colhead{RA} & \colhead{Dec} & \colhead{z} &\colhead{Type} &\colhead{Galaxy} &\colhead{W2}& \colhead{W1-W2} &\colhead{ZTF }&\colhead{WISE }&\colhead{VLASS 3GHz}&\colhead{RACS 1GHz}\\[-0.3cm]
\colhead{BAT ID} & \colhead{(hms)} & \colhead{(dms)} & \colhead{}& \colhead{}& \colhead{Morphology} &\colhead{(mag)}& \colhead{(mag)} &\colhead{variance}&\colhead{$\chi^2$/dof}&\colhead{(mJy)}&\colhead{(mJy)}}
\startdata
2&00:01:46.08&-76:57:14.40&$0.058$&Sy1.5&merger&$10.6$&$0.84$&$-$&904.52&-&$6.48\pm1.14$\\
3&00:02:26.42&$+$03:21:06.84&$0.025$&Sy1.2&disk-spiral&$10.27$&$0.61$&$0.07$&846.56&ND&$8.77\pm1.68$\\
6&00:06:19.54&$+$20:12:10.80&$0.0262$&Sy1.2&point-like&$7.84$&$0.89$&$0.06$&1540.96&$4.97\pm0.31$&-\\
14&00:26:40.68&-53:09:47.88&$0.062$&Sy1.5&smooth&$10.75$&$0.67$&$-$&2109.13&-&$1.44\pm0.7$\\
22&00:36:20.95&$+$45:39:53.64&$0.0477$&Sy1.2&disk-spiral&$9.89$&$0.76$&$0.07$&204.51&ND&-\\
29&00:43:01.90&$+$30:17:19.68&$0.052$&Sy1.9&smooth&$10.46$&$0.85$&$0.61$&1246.27&$1.4\pm0.32$&-\\
36&00:51:54.77&$+$17:25:58.44&$0.0649$&Sy1.2&smooth&$10.8$&$0.96$&$0.06$&22.58&ND&-\\
39&00:54:52.13&$+$25:25:39.00&$0.155$&Sy1.5&smooth&$10.19$&$1.01$&$0.1$&581.49&$1.23\pm0.29$&$3.08\pm0.91$\\
51&01:05:38.81&-14:16:13.44&$0.0664$&Sy1.5&other/unc.&$10.21$&$1.12$&$0.05$&777.73&$3.81\pm0.24$&$8.73\pm1.34$\\
55&01:07:39.65&-11:39:11.16&$0.0466$&Sy1.8&edge-on&$10.58$&$0.96$&$0.03$&17.91&$5.67\pm0.85$&$10.57\pm1.68$\\
60&01:13:51.05&$+$13:16:18.48&$0.049$&Sy1.5&merger&$9.05$&$0.95$&$0.03$&106.93&$5.9\pm0.28$&$15.78\pm2.08$\\
65&01:16:31.15&-12:36:16.92&$0.1425$&Sy1.8&smooth&$11.35$&$1.05$&$-$&134.64&$3.6\pm0.25$&$5.59\pm1.24$\\
72&01:23:54.36&-35:03:55.44&$0.019$&Sy1.9&merger&$7.92$&$0.95$&$-$&671.11&$5.96\pm0.28$&$19.02\pm1.98$\\
73&01:23:45.77&-58:48:20.88&$0.047$&Sy1.2&merger&$7.94$&$0.98$&$-$&327.31&-&$5.84\pm1.02$\\
75&01:25:55.94&$+$35:10:36.84&$0.3119$&Sy1&smooth&$9.82$&$1.0$&$0.12$&208.14&$18.79\pm0.23$&-\\
78&01:28:06.72&-18:48:30.96&$0.046$&Sy1.5&disk-spiral&$9.92$&$0.87$&$0.02$&3940.59&$1.92\pm0.39$&$4.06\pm1.02$\\
80&01:29:07.66&-60:38:42.00&$0.2036$&Sy1.8&merger&$12.13$&$0.94$&$-$&86.33&-&$13.85\pm1.59$\\
85&01:34:45.62&-04:30:13.32&$0.079$&Sy1.5&smooth&$10.9$&$0.76$&$0.59$&1441.62&$1.21\pm0.3$&-\\
88&01:39:24.00&$+$29:24:07.20&$0.072$&Sy1.5&merger&$11.12$&$0.42$&$0.11$&84.4&$2.96\pm0.32$&-\\
89&01:40:26.81&-53:19:39.36&$0.0716$&Sy1.5&smooth&$10.86$&$1.0$&$-$&3373.91&-&$11.28\pm1.35$\\
92&01:48:59.69&$+$21:45:33.84&$0.0691$&Sy1.2&smooth&$12.29$&$0.2$&$0.49$&62.6&ND&-\\
98&01:55:24.96&$+$02:28:16.68&$0.0828$&Sy1&smooth&$10.44$&$0.87$&$0.52$&1073.76&$1.55\pm0.28$&-\\
99&01:57:10.94&$+$47:15:59.04&$0.048$&Sy1.2&disk-spiral&$10.94$&$0.59$&$1.59$&565.53&ND&-\\
\enddata
\vspace{0.1cm}
\tablecomments{Properties of the 272 BASS AGN that were not classified as DPEs and served as a non-DPE broad-line AGN sample. Col. 1: BAT ID number from \citet{Koss2022} Cols. 2-3: RA and DEC of BAT counterpart from \citet{Koss2022}. Col 4: spectroscopic redshift from \citet{Koss2022}. Col. 5: Seyfert classification from \citet{Oh2022}. Col. 6: galaxy morphology classification from \citet{ParraTello2025}. Cols. 7 and 8: median W2 magnitude and median W1-W2 color across the NEOWISE light curves. Col. 9: excess variance from the g-band ZTF light curves. Col. 10: $\chi^2\text{/dof}$ of the WISE W2 light curves. Col. 11: 2-4 GHz radio flux from VLASS for epoch 1 (2017-2018), with 2.5$"$ beam, where ND indicates radio non-detection and dash indicates that the source was not within the surveyed region. Col. 12: 1GHz radio flux from ASKAP-RACS, with 15$"$ resolution.}
\end{deluxetable*}

\software{
Astropy \citep{astropy:2013,astropy:2018,astropy:2022}, ppxf \citep{Cappellari2017ImprovingFunctions}, SciPy \citep{2020SciPy-NMeth} 
}

\bibliography{main,references,IMBHs}{}
\bibliographystyle{aasjournal}

\end{document}